\RequirePackage{ifpdf}
\ifpdf % We are running pdfTeX in pdf mode
\documentclass[pdftex]{sigma}
\else
\documentclass{sigma}
\fi

\numberwithin{equation}{section}

\begin{document}

\allowdisplaybreaks

\renewcommand{\thefootnote}{$\star$}

\renewcommand{\PaperNumber}{004}

\FirstPageHeading

\ShortArticleName{Classical Particle in Presence of Magnetic Field}

\ArticleName{Classical Particle in Presence of Magnetic Field,\\  Hyperbolic
 Lobachevsky\\ and Spherical Riemann Models\footnote{This
paper is a contribution to the Proceedings of the Eighth
International Conference ``Symmetry in Nonlinear Mathematical
Physics'' (June 21--27, 2009, Kyiv, Ukraine). The full collection
is available at
\href{http://www.emis.de/journals/SIGMA/symmetry2009.html}{http://www.emis.de/journals/SIGMA/symmetry2009.html}}}

\Author{V.V. KUDRYASHOV, Yu.A. KUROCHKIN, E.M. OVSIYUK and V.M. RED'KOV}

\AuthorNameForHeading{V.V. Kudryashov, Yu.A. Kurochkin, E.M. Ovsiyuk and V.M. Red'kov}

\Address{Institute of Physics,  National Academy of Sciences of Belarus, Minsk, Belarus}

\Email{\href{mailto:kudryash@dragon.bas-net.by}{kudryash@dragon.bas-net.by}, \href{mailto:y.kurochkin@ifanbel.bas-net.by}{y.kurochkin@ifanbel.bas-net.by},\\
\hspace*{13.5mm}\href{mailto:mozlena@tut.by}{mozlena@tut.by}, \href{mailto:redkov@dragon.bas-net.by}{redkov@dragon.bas-net.by}}

\ArticleDates{Received July 20, 2009, in f\/inal form December 29, 2009;  Published online January 10, 2010}

\Abstract{Motion of a classical  particle in 3-dimensional Lobachevsky  and
Riemann spaces  is studied in the presence of an external magnetic
f\/ield which is analogous to a constant uniform magnetic
 f\/ield in Euclidean space.
 In both cases three integrals of motions are constructed and equations of motion
 are solved exactly in the special cylindrical coordinates on the base of the method of separation
 of variables. In Lobachevsky space there exist trajectories of two types, f\/inite and inf\/inite in radial variable,
in Riemann space all  motions are f\/inite and periodical.
 The invariance   of the uniform magnetic f\/ield in tensor description and gauge invariance of corresponding
 4-potential description is demonstrated explicitly.
The role  of the symmetry is  clarif\/ied in classif\/ication of all
possible solutions, based on the geometric  symmetry
group, $SO(3,1)$ and $SO(4)$ respectively.}


\Classification{37J35; 70G60; 70H06; 74H05}

%\noindent CONTENT
%
%\vspace{2mm}
%
%\noindent PART I.  PARTICLE IN THE MODEL $H_{3}$
%
%
%\vspace{2mm}
%
%\noindent 1. Introduction
%
%
%\noindent 2.  Newton second law in Lobachevsky space
%
%
%\noindent 3. Particle in a uniform magnetic f\/ield, hyperbolic model $H_{3}$
%\noindent 4.  Simplest solutions in Lobachevsky model
%
%\noindent 5. Particle in a magnetic f\/ield and Lagrange formalism in $H_{3}$
%
%\noindent 6. Possible solutions in $H_{3}$, radial f\/inite and inf\/inite motions
%
%\noindent 7. Trajectory equation in the form $F(r,z)=0$, model $H_{3}$
%
%\noindent 8. Trajectory equation  $F(r,\phi)=0$,  the role of Lorentz $SO(3,1)$ transversal shifts
%\noindent   in Lobachevsky space
%
 %\noindent 9.  Lorentzian  shifts   and  symmetry  of a magnetic f\/ield  in $H_{3}$
%
%\vspace{2mm}
%
%\noindent PART II. PARTICLE IN THE MODEL $S_{3}$
%\vspace{2mm}
%
%\noindent 10. Particle in a magnetic f\/ield, spherical Riemann model $S_{3}$
%
%\noindent 11. Simplest solutions in spherical space
%
%\noindent 12. Particle in a magnetic f\/ield and Lagrange formalism in   $S_{3}$
%
%\noindent 13.  All  trajectories  and $SO(4)$ symmetry of the space $S_{3}$
%
%\noindent 14.  Space  shifts   and gauge symmetry  of the magnetic f\/ield  in $H_{3}$
%
%\noindent 15. Extension to relativistic case
%
%\noindent 16. Discussion
%
%\noindent 17. Acknowledgements
%
%\noindent References

\renewcommand{\thefootnote}{\arabic{footnote}}
\setcounter{footnote}{0}

\section{Introduction}

In the paper exact solutions for classical problem of a
particle in a magnetic f\/ield
 on the background
of hyperbolic Lobachevsky $H_{3}$ and spherical Riemann $S_{3}$
space models will be  constructed explicitly.

The grounds to examine these problems are as follows: these both
are extensions  for a well-known  problem in
theoretical physics~-- a particle in  a uniform magnetic f\/ield \cite{Landau-Lifshitz-2};
they  can be  used
to describe behavior of charged particles in macroscopic magnetic
f\/ield in the context of  astrophysics. The form of the
 magnetic f\/ield   in the models $H_{3}$ and $S_{3}$
was introduced earlier  in
 \cite{BRK-1,BRK-2,BRK-3} where   the
quantum-mechanical variant  (for the Shr\"{o}dinger equation) of the
problem had been solved as well and generalized  formulas for
Landau levels  \cite{Landau-1930, Landau-Lifshitz-2, Landau-Lifshitz-3}
had been produced. A part of results of the
 paper was presented in the talk given in \cite{KKOR-2009}.

Previously, the main attention was given to Landau problem in 2-dimensional case:
many important  mathematical and physical results were obtained, see in
\cite{Klauder-Onofri-et-1992,Cappelli-Dunne-Trugenberger-Zemba-1993,
Cappelli-Trugenberger-Zemba-1993,
Drukker et-2003,
Dunne-1992,
Klauder-Onofri-1989,
Negro et-2001,
Onofri-2001}.
Comprehensive  discussion of the general problem of integrability
of classical and quantum systems in Lobachevsky and Riemann 3D and 2A
models  see in  \cite{Carinera et al-2005, Gadella-Negro-Pronko-Santander-2007, Herranz-Ballesteros-2006}
 and references therein.
 It is known that  2D and 3D systems
exhibit  properties  which are radically dif\/ferent.
Our treatment will concern only a 3-dimensional case.

\section{Newton second law in Lobachevsky space}

Motion of a classical particle in external electromagnetic and
gravitational f\/ields is described by the known equation
\cite{Landau-Lifshitz-2}
\begin{gather}
mc^{2} \left(   {d^{2} x^{\alpha}  \over d s^{2}  } +
\Gamma^{\alpha}_{\;\;\beta \sigma }   {dx^{\beta} \over d s } {d
x^{\sigma} \over d s} \right) = e F^{\alpha \rho}  U_{\rho} ,
\label{1.1}
\end{gather}
where Christof\/fel symbols are determined by metrical
structure of a space-time (the signature  $+\,-\,-\,-$ is used). In (\ref{1.1}) it is useful to perform
$(3+1)$-splitting
\begin{gather}
mc^{2}
    {d^{2} x^{0}  \over d s^{2}  }  =  e    \big( F^{0 1}  U_{1}  +
  F^{0 2}  U_{2}  + F^{0 3}  U_{3} \big)   ,
\label{1.5a}
\\
mc^{2} \left(    {d^{2} x^{1}  \over d s^{2}  } + \Gamma^{1}_{\;\;jk }
 {dx^{j} \over d s } {d x^{k} \over d s} \right) = e F^{1 0}  U_{0} +
e F^{1 2}  U_{2} +  e F^{1 3}  U_{3}     , \label{1.5aa}
\\
mc^{2} \left(    {d^{2} x^{2}  \over d s^{2}  } + \Gamma^{2}_{\;\;jk }
  {dx^{j} \over d s } {d x^{k} \over d s} \right) = e F^{2 0}  U_{0} +
e F^{2 1}  U_{1}  + + e F^{2 3}  U_{3}   , \label{1.5aaa}
\\
mc^{2} \left(    {d^{2} x^{3}  \over d s^{2}  } + \Gamma^{3}_{\;\;jk }
  {dx^{j} \over d s } {d x^{k} \over d s} \right) = e F^{3 0}  U_{0} +
e F^{3 1}  U_{1} + e F^{3 2}  U_{2}     . \label{1.5b}
\end{gather}
 In  (\ref{1.5a})--(\ref{1.5b})  usual SI units  are used, so  the Christof\/fel symbols $\Gamma^{i}_{jk}$ are  measured
in $\mbox{(meter)}^{-1}$.  With conventional notation
\cite{Landau-Lifshitz-2}
\begin{gather*}
\big(F^{\alpha \beta} \big) =  \left | \begin{array}{rrrr}
0     & -E^{1}  & -E^{2}  & - E^{3} \\
E^{1} &    0    & -cB^{3} & cB^{2} \\
E^{2} & -cB^{3} &    0    & -cB^{1} \\
E^{3} & -cB^{2} & cB^{1}  &  0
\end{array} \right |  , \\
U^{\alpha}  = {dt \over ds}     {d \over dt} \big(dx^{0}, dx^{i} \big)
= {1 \over \sqrt{1 - V^{2} /c^{2}}}   \left (1 ,   {V^{i} \over
c } \right)  , \\
V^{i} =   {dx^{i} \over  d t}    , \qquad  V^{2} = - g_{ki} (x)
V^{k} V^{i}   , \qquad { d x^{0} \over ds} =  \left( { c dt   \sqrt{1
- V^{2} / c^{2}} \over c dt} \right)^{-1} = {1 \over \sqrt{1 - V^{2} /
c^{2}} }   ,
\end{gather*}
equations  (\ref{1.5a})--(\ref{1.5b})  give
\begin{gather*}
 {d  \over    dt}
 \left( { mc^{2} \over \sqrt{1 - V^{2} / c^{2}} } \right) =e     \big[   -  g_{ik} (x)   E^{i} V^{k}    \big],
%\label{1.6a}
\\
{d \over dt} {V^{1} \over  \sqrt{1 - V^{2} /c^{2}}} + {1 \over
\sqrt{1 - V^{2} /c^{2}}}   \Gamma^{1}_{\;\;jk }    V^{j} V^{k}   = {e \over  m}  E^{1}   - {e \over m}
\big( V_{2} B^{3} -   V_{3} B^{2}\big)     , %\label{1.6aa}
\\
 {d \over dt} {V^{2} \over  \sqrt{1 - V^{2} /c^{2}}} +
{1 \over \sqrt{1 - V^{2} /c^{2}}}   \Gamma^{2}_{\;\;jk }
V^{j} V^{k}   = {e \over  m}  E^{2}
- {e \over  m}  \big( V_{3} B^{1} -   V_{1} B^{3}\big)       , %\label{1.6aaa}
\\
 {d \over dt} {V^{3} \over  \sqrt{1 - V^{2} /c^{2}}} +
{1 \over \sqrt{1 - V^{2} /c^{2}}}   \Gamma^{3}_{\;\;jk }
V^{j} V^{k}  = {e \over  m}  E^{3}
- {e \over  m}\big( V_{1} B^{2} -   V_{2} B^{1}\big)      .
%\label{1.6b}
\end{gather*}

Firstly, we will be interested in non-relativistic
case\footnote{Extension to the relativistic case
will be performed in the end of the paper.}, when  all
equations become simpler
\begin{gather}
 {d  \over  dt}    { m V^{2} \over 2 }  =e     \big(   -  g_{ik}    E^{i} V^{k}     \big)  ,
\label{1.7a}
\\
{d \over dt} V^{1}+
 \Gamma^{1}_{\;\;jk }    V^{j} V^{k}   =
{e \over  m}   E^{1}   - {e \over  m}  \big( V_{2} B^{3} -   V_{3}
B^{2}\big)    , \label{1.7aa}
\\
 {d \over dt} V^{2}  +
 \Gamma^{2}_{\;\;jk }   V^{j} V^{k}  =
{e \over  m}   E^{2}   - {e \over  m} \big( V_{3} B^{1} -   V_{1}
B^{3}\big)       , \label{1.7aaa}
\\
{d \over dt} V^{3} + \Gamma^{3}_{\;\;jk }   V^{j}V^{k} = {e
\over  m}   E^{3}   - {e \over  m}\big( V_{1} B^{2} - V_{2} B^{1}\big)
  . \label{1.7b}
\end{gather}

\section[Particle in a uniform magnetic field, hyperbolic model $H_{3}$]{Particle in a uniform magnetic f\/ield, hyperbolic model $\boldsymbol{H_{3}}$}\label{section3}

Let us start with the known 4-vector potential of a uniform
magnetic f\/ield in f\/lat space  \cite{Landau-Lifshitz-2}
\begin{gather}
{\bf A} = {1 \over 2}   c {\bf B} \times {\bf r}, \qquad {\bf B}
= (0,0,B)   , \qquad {A^{a} \over c}  =  {B  \over 2}   (0 ;
-r\sin \phi, r\cos \phi, 0)  . \label{2.1a}
\end{gather}
From  (\ref{2.1a}), after transformation to cylindric coordinates we obtain
\begin{gather*}
A_{t} = 0  , \qquad A_{r} = 0    , \qquad A_{z} = 0   , \qquad
A_{\phi} =  -{cBr^{2} \over 2}  . %\label{2.1c}
\end{gather*}
The only non-vanishing constituent of the electromagnetic
tensor reads
\begin{gather*}
F_{\phi r} =
\partial_{\phi}A_{r} - \partial_{r} A_{\phi} = c B r   ,
%\label{2.1d}
\end{gather*}
which satisf\/ies the Maxwell equations
\begin{gather*}
 {1 \over \sqrt{-g}}
{\partial \over \partial x^{\alpha} }\sqrt{-g} F^{\alpha \beta} =
0 \   \Longrightarrow \  {1 \over r } {\partial \over
\partial r } r  F^{\phi  r} = {1 \over r } {\partial \over
\partial r } r   \left( {1 \over r^{2}} \right) cBr  \equiv 0   .
\end{gather*}

Now we are to extend the concept of a uniform magnetic f\/ield to
the Lobachevsky model~$H_{3}$. Thirty four orthogonal coordinate systems in
this space were found  by Olevsky~\cite{Olevsky}. An idea is to
select  among them some curved analogue for cylindric coordinates
and determine  with their  help an appropriate solution to Maxwell
equations in Lobachevsky space. In~\cite{Olevsky},  under  the number~XI  we see  the following coordinates
\begin{gather}
dS^{2} = c^{2} dt^{2} - \rho^{2}   \big[  \cosh^{2} z   \big( d
r^{2} + \sinh^{2} r   d \phi^{2} \big) + dz^{2}\big] ,
\nonumber
\\
z \in ( - \infty , + \infty )   , \qquad r \in [0, +\infty )   ,
\qquad \phi \in [0, 2 \pi ]   , \nonumber
\\
u_{1} = \cosh z  \sinh r \cos \phi  , \qquad u_{2}
= \cosh  z \sinh r \sin \phi   ,\nonumber\\
 u_{3} =
\sinh z  , \qquad   u_{0} = \cosh  z \cosh r
, \nonumber
\\
u_{0}^{2} - u_{1}^{2} - u_{2}^{2} - u_{3}^{3} = 1  , \qquad
u_{0} = + \sqrt{1 + {\bf u}^{2} }   , \label{2.2}
\end{gather}
the curvature radius   $\rho$ is taken  as a unit
length. In the limit $\rho \to \infty$ the coordinates
(\ref{2.2}) reduce to ordinary cylindric ones in the f\/lat space. By
def\/inition, the uniform magnetic f\/ield in the Lobachevsky space is
given by 4-potential of the form
\begin{gather*}
 A_{\phi} = -2cB \rho^{2}   \sinh^{2} {r \over 2}  = - cB\rho^{2}  (\cosh  r -1 )  .
%\label{2.3a}
\end{gather*}
It behaves properly in the limit $\rho \longrightarrow
\infty$, besides  it corresponds to an electromagnetic
tensor which evidently satisf\/ies  the  Maxwell equations in $H_{3}$
\begin{gather*}
F_{\phi r} =
 - {1 \over \rho }  \partial_{r} A_{\phi} =  cB  \rho   \sinh  r   ,
\nonumber\\
 {1 \over
\cosh^{2} z   \sinh  r } {\partial \over
\partial r }  \cosh^{2} z   \sinh  r  \left( {1 \over
\cosh^{4}z    \sinh^{2}{r} } \right)B   \sinh  r  \equiv 0
  . %\label{2.3b}
\end{gather*}

In the absence of an electric f\/ield, the non-relativistic equations
(\ref{1.7aa})--(\ref{1.7b})  read
\begin{gather}
 {d V^{r}  \over d t   } +
\Gamma^{r}_{\;\;jk }   V^{j}  V^{k}  = {e \over m}   F^{r \phi}
 V_{\phi}   ,  \nonumber
\\
{d  V^{\phi}  \over d t  } + \Gamma^{\phi}_{\;\;jk }   V^{j}
V^{k}  =
 {e \over m}     F^{\phi r}  V_{r}    ,
\nonumber\\
 {d  V^{z}  \over d t  } + \Gamma^{z}_{\;\;jk }   V^{j}
V^{k} =  0     , \label{2.5}
\end{gather}
where  the Christof\/fel symbols are
\begin{gather*}
\Gamma^{r}_{\;\;jk } = \left | \begin{array}{ccc}
0 & 0 & \tanh z \\
0 & - \sinh  r   \cosh  r & 0 \\
\tanh\;z  & 0 & 0
\end{array} \right |   , \qquad
\Gamma^{\phi}_{\;\;jk } = \left | \begin{array}{ccc}
0 &  \coth  r  & 0\\
\coth  r  & 0 & \tanh  z \\
0 & \tanh  z  & 0
\end{array} \right |  ,
\nonumber
\\
\Gamma^{z}_{\;\;jk } = \left | \begin{array}{ccc}
-\cosh  z  \sinh  z  &  0  & 0\\
0  &  -\sinh  z   \cosh  z   \sinh^{2} r & 0 \\
0 & 0 & 0
\end{array} \right |  .
%\label{2.6}
\nonumber
\end{gather*}

It makes sense  to recall  a similar problem in f\/lat space, here
the Christof\/fel symbols are simpler
\begin{gather*}
\Gamma^{r}_{\;\;jk } = \left | \begin{array}{ccc}
0 & 0 & 0 \\
0 & - r  & 0 \\
0 & 0 & 0
\end{array} \right |   , \qquad
\Gamma^{\phi}_{\;\;jk } = \left | \begin{array}{ccc}
0 &  r^{-1} & 0 \\
r^{-1}  & 0 &  0 \\
0 & 0  & 0
\end{array} \right |, \qquad
\Gamma^{z}_{\;\;jk } = 0  % \label{2.7}
\end{gather*}
 and equations of motion  read
in space  $E_{3}$,
\begin{gather}
 {d V^{r}  \over d t   }  - r V^{\phi} V^{\phi}  = {e \over m} B  r V^{\phi}     ,
\qquad {d  V^{\phi}  \over d t  } +
 {2 \over r}  V^{r} V^{\phi}  =-
 q  {B \over r} V^{r}  , \qquad
{d  V^{z}  \over d t  }  =  0    . \label{2.10}
\end{gather}
In fact, a simplest solution of these equations is
well-known: the particle moves on  a cylindric surface oriented
along the axis $z$ according to the law $\phi (t) = \omega t +
\phi_{0}$, correspondingly  equations~(\ref{2.10}) take the form
\begin{gather*}
{d^{2} r  \over d t^{2}   }  = r  \omega   \left(\omega + {e \over
m}  B \right)      , \qquad
 {dr \over dt}  {1 \over r}   \left(  2    \omega   +  {e \over m}   B \right) =0    \ \Longrightarrow  \   {dr \over dt}=0      ,
\end{gather*}
 and  the simplest  solution is given by
\begin{gather*}
\omega = - {e B \over m}   ,  \qquad \phi (t) = \omega t +
\phi_{0}   , \qquad r(t) =r_{0}    , \qquad z(t) = z (t) = z_{0}
+ V^{z}_{0}   t  . %\label{2.11}
\end{gather*}

Now, let us turn to the problem  in Lobachevsky model -- equations~(\ref{2.5}) give
\begin{gather}
   {d V^{r}  \over d t   } +
2   \tanh  z   V^{r}  V^{z} - \sinh  r   \cosh
r   V^{\phi }  V^{\phi} =
  B   { \sinh  r    \over \cosh^{2} z }  V^{\phi}      ,
\nonumber
\\
{d  V^{\phi}  \over d t  } + 2   \coth  r  V^{\phi }
V^{r}  +   2   \tanh  z    V^{\phi } V^{z} =
 - B \;  {1   \over \cosh^{2} z   \sinh\;  r}     V^{r}      ,
\nonumber
\\
{d  V^{z}  \over d t  }  -\sinh  z  \cosh  z     V^{r}
V^{r} -\sinh  z   \cosh  z   \sinh^{2} r
V^{\phi}  V^{\phi}=  0     . \label{2.12}
\end{gather}
It should be stressed that in (\ref{2.12}) all
quantities (coordinates $t$, $r$, $\phi$, $z $ as well ) are {\it
dimensionless}\footnote{Bellow in the paper all relationships are
written in that dimensionless  form.}; in particular, symbol  $B$
stands for a special combination of magnetic f\/ield amplitude,
charge, mass, light velocity, and  curvature radius
\begin{gather*}
B  \  \Longleftrightarrow  \  {e \over m}   { \rho B \over  c
}   , \qquad t  \   \Longleftrightarrow   \   {c t \over \rho
}  , \qquad
 r  \   \Longleftrightarrow \   {r \over \rho}   ,
\qquad z  \   \Longleftrightarrow \   {z \over \rho}   .
%\label{2.13}
\end{gather*}

\section{Particular solutions in Lobachevsky model}

Let us construct simple solutions when  imposing  the following
constrain $ r = r_{0}  = \mbox{const}  $,
 equations (\ref{2.12}) give
\begin{gather}
    V^{\phi }  = -   { B  \over \cosh  r_{0}}   { 1  \over \cosh^{2} z }       ,\nonumber\\
 {d    \over d t  }  \left( -   {B  \over \cosh\; r_{0}} \; {
1  \over \cosh^{2} z } \right) +
    2   \tanh  z   \left(  -   { B  \over \cosh  r_{0}}  { 1  \over \cosh^{2} z }\right)  V^{z} = 0    ,
\nonumber
\\
{d  V^{z}  \over d t  }  =  \big(   \tanh^{2} r_{0}     B^{2}  \big)
 {  \sinh  z  \over \cosh^{3} z }   . \label{3.1}
\end{gather}

\noindent The second equation is an identity $0 \equiv 0$.
With the  notation
\[
\alpha = -    B  /  \cosh  r_{0} ,  \qquad A = \big(
\tanh^{2} r_{0}    B^{2}  \big) > 0  ,
\]
 two remaining equations in  (\ref{3.1}) read
\begin{gather}
    {d \phi \over dt}   =  { \alpha  \over \cosh^{2} z }       , \qquad
{d  V^{z}  \over d t  }  =  A  {  \sinh  z  \over
\cosh^{3} z }    . \label{3.2}
\end{gather}
When $B>0 $, the angular velocity $d\phi / dt< 0$; and
when  $B<0 $, the angular velocity $d\phi / dt > 0$. Second
equation in  (\ref{3.2})  means  that there exists ef\/fective
repulsion  to both sides from the center  $z=0$. One can  resolve
the second equation
\begin{gather}
d  ( V^{z})^{2}  =  A     d \left( -{1 \over \cosh^{2} z}  \right)
\qquad  \Longrightarrow \qquad
 \left({dz \over dt}\right)^{2} = \epsilon  -{A \over \cosh^{2} z }    .
\label{3.3}
\end{gather}

Below we will see that the constant $\epsilon$  can be related to a squared velocity $V^{2} / c^{2}$  or
dif\/ferently to the integral of motion~-- the energy of the
non-relativistic particle\footnote{Relationship (\ref{3.3}) in the
limit of f\/lat space will read $(dz / dt)^{2} =  \epsilon  - A $
which means that $A$ corresponds to a transversal squared velocity
$V^{2}_{\bot } / c^{2}$. Also, equation~(\ref{3.3}) points out that
in  Lobachevsky model the transversal motion should vanish (to be
frozen) when $z \rightarrow \pm \infty$.}.
First, let  $A \neq  \epsilon$
\begin{gather*}
\pm   {  d   \sinh   z  \over \sqrt{ \epsilon  (1 +
\sinh^{2} z )   - A }  } = dt  . %\label{3.4}
\end{gather*}
Meaning of the signs  $\pm$  is evident:   it corresponds
to a motion along axis~$z$ in opposite directions. Further we get
\begin{gather*}
{\rm I.} \quad  \underline{\epsilon > A } , \qquad  z \in ( -
\infty, + \infty )   , \qquad \pm {1 \over \sqrt{\epsilon}} \,
{\rm arcsinh}
 \left(   \sqrt{{\epsilon \over  \epsilon - A}}  \sinh  z   \right) = t - t_{0}   ;
\nonumber
\\
{\rm II.} \quad  \underline{\epsilon < A  }, \qquad
\sinh^{2} z > {A - \epsilon \over  \epsilon }   , \qquad \pm
{1 \over \sqrt{\epsilon}} \, {\rm arccosh}
 \left(   \sqrt{{\epsilon \over  A - \epsilon}}  \sinh  z  \right) = t - t_{0}   ,
 %\label{3.4a}
 \end{gather*}
 or
\begin{gather*}
{\rm I.} \quad  \underline{\epsilon > A} , \qquad  z \in ( -
\infty, + \infty ) , \qquad
 \sinh z (t) = \pm  {
\sqrt{\epsilon -A} \over \sqrt{\epsilon}} \sinh
\sqrt{\epsilon}  (t - t_{0})  \nonumber
\\
{\rm II.} \quad  \underline{\epsilon < A  }, \qquad
\sinh^{2} z > {A - \epsilon \over  \epsilon }   , \qquad
 \sinh  z (t) = \pm   {
\sqrt{A - \epsilon } \over \sqrt{\epsilon}}   \cosh
\sqrt{\epsilon}   (t - t_{0})    . %\label{3.5}
\end{gather*}

For motions of the type~I,  trajectories run through  $z=0$; for
motions  of the type~II,  the particle is repulsed  from the
center $z=0$ at the points  $\sinh  z_{0} = \pm \sqrt{A /
\epsilon -1 }  $. Existence of these two dif\/ferent regimes  of
motion along the  axis~$z$ correlates with the mentioned ef\/fective
repulsion along the axis~$z$.

Now let us consider the case  $\epsilon = A$
\begin{gather}
 \left({dz \over dt}\right)^{2} = \epsilon    \tanh^{2} z    .
\label{3.6}
\end{gather}
We immediately see a trivial solution
\begin{gather*}
z (t) = 0 \   \Longrightarrow \   \phi (t) = \phi _{0} +
\alpha   t   , \qquad \alpha = -   { B  \over \cosh  r_{0}}
 ; %\label{3.7}
\end{gather*}
it corresponds to rotation of the particle along the
circle $r=r_{0}$  in the absence of any motion along the axis~$z$. Also
there are  non-trivial solutions to equation (\ref{3.6})
\begin{gather}
{d   \sinh   z \over  \sinh  z} =   \pm \sqrt{\epsilon}
  dt   ;
\label{3.8}
\end{gather}
here we have two dif\/ferent ones  depending on  sign
 $(+)$  or $(-)$. Continuous solutions of (\ref{3.8}) exist only for
$z>0$  and    $z <0$ with  dif\/ferent and peculiar properties.
In the case of   sign ($+$) we have
\begin{gather}
\sinh  z = \sinh z_{0}  e^{+ \sqrt{\epsilon} t}   ,
\qquad  (t=0 , \  z = z_{0} \neq 0)   ; \label{3.9a}
\end{gather}
at any positive initial  $z_{0}>0$  the particle goes to
$+\infty$; and at any negative initial  $z_{0}<0 $ the particle
goes to $-\infty$.
In the case of  sign ($-$), we get quite dif\/ferent behavior
\begin{gather}
\sinh  z = \sinh z_{0}   e^{- \sqrt{\epsilon} t}   ,
\qquad  (t=0 , \  z = z_{0} \neq 0)   ; \label{3.9b}
\end{gather}
at any positive initial  $z_{0}>0$  the particle moves
to~$z=0$ during inf\/inite time~$t$; at any negative initial
$z_{0}<0 $ the particle moves to  $z=0$ during inf\/inite time~$t$.

Now we are to turn to the f\/irst equation in   (\ref{3.2}) and f\/ind
$\phi (t)$
\begin{gather}
\epsilon > A  , \qquad \phi - \phi _{0}    =   {  \alpha  \over
\sqrt{ A} } \,  \mbox{arccoth}\, \left( \sqrt{{A \over \epsilon}} \;
\tanh \; \sqrt{\epsilon}  t \right), \nonumber
\\
\epsilon < A   , \qquad \phi - \phi _{0}    =   {  \alpha  \over
\sqrt{ A} } \,  \mbox{arccoth}\, \left( \sqrt{{\epsilon \over A}}
\tanh   \sqrt{\epsilon}  t \right) . \label{3.10a}
\end{gather}
One may note that when  $t \rightarrow + \infty$ we
obtain  a  f\/inite value for the rotation angle
\begin{gather*}
\epsilon > A   , \qquad \left. (\phi - \phi _{0}) \right |_{t
\rightarrow \infty}      =   {  \alpha  \over \sqrt{ A} } \,
 \mbox{arccoth}\,  \sqrt{{A \over \epsilon}}    ,
 \nonumber
 \\
 \epsilon < A   , \qquad
\left. (\phi - \phi _{0}) \right |_{t \rightarrow \infty}      =
{  \alpha  \over \sqrt{ A} } \,
 \mbox{arccoth}\,  \sqrt{{\epsilon \over A }}    .
%\label{3.10b}
\end{gather*}

In the same manner one should f\/ind  $\phi (t)$ in case
(\ref{3.9a})
\begin{gather*}
A = \epsilon  , \qquad \sinh  z = \sinh z_{0}   e^{+
\sqrt{\epsilon} t}   , \nonumber
\\
\phi - \phi _{0}    =  \alpha    \int  {   dt  \over
\cosh^{2} z }  = \alpha   \int {d t \over 1 + \sinh^{2}
z_{0}   e^{+ 2\sqrt{\epsilon} t}  }   , \qquad x = \sinh^{2}
z_{0}  e^{+ 2\sqrt{\epsilon} t}   , \nonumber
\\
\left. \phi - \phi _{0} = { \alpha \over 2 \sqrt{\epsilon}}   \ln
{x \over x+1} \right |_{t = 0}^{t} = { \alpha \over 2
\sqrt{\epsilon}}   \left [    \ln { \sinh^{2}z_{0} \over
\sinh^{2}z_{0} + e^{-2 \sqrt{\epsilon}   t}} - \ln {
\sinh^{2}z_{0} \over  \sinh^{2}z_{0} + 1 }    \right ]; \nonumber
\end{gather*}
so that
\begin{gather*}
\phi - \phi _{0} = { \alpha \over 2 \sqrt{\epsilon}}  = \ln   {
\sinh^{2}z_{0}  +1 \over  \sinh^{2}z_{0} + e^{-2
\sqrt{\epsilon}   t}}   , \qquad z_{0} \neq 0  . %\label{3.11a}
\end{gather*}
Again a peculiarity in  the limit  $t \rightarrow +
\infty $  may be noted
\begin{gather*}
t \rightarrow + \infty   , \qquad \phi - \phi _{0}  = { \alpha
\over 2 \sqrt{\epsilon}} \ln     { \sinh^{2}z_{0}  +1 \over
\sinh^{2}z_{0} }   , \qquad z_{0} \neq 0   .% \label{3.11b}
\end{gather*}

And the function  $\phi (t)$  in case  (\ref{3.9b})
\begin{gather*}
A = \epsilon  , \qquad \sinh  z = \sinh z_{0}   e^{-
\sqrt{\epsilon} t}   , \nonumber
\\
\phi - \phi _{0}    =  \alpha    \int  {   dt  \over
\cosh^{2} z }  = \alpha   \int {d t \over 1 + \sinh^{2}
z_{0} \; e^{- 2\sqrt{\epsilon} t}  }   , \qquad x = \sinh^{2}
z_{0} \; e^{- 2\sqrt{\epsilon} t}   , \nonumber
\\
\left. \phi - \phi _{0} = - { \alpha \over 2 \sqrt{\epsilon}}
\ln {x \over x+1} \right |_{t = 0}^{t} =
 -
{ \alpha \over 2 \sqrt{\epsilon}}   \left [    \ln {
\sinh^{2}z_{0} \over  \sinh^{2}z_{0} + e^{+2
\sqrt{\epsilon}   t}} - \ln { \sinh^{2}z_{0} \over
\sinh^{2}z_{0} + 1 }    \right ], \nonumber
\end{gather*}
so that
\begin{gather*}
\phi - \phi _{0} =  - { \alpha \over 2 \sqrt{\epsilon}}     \ln
  { \sinh^{2}z_{0}  +1 \over  \sinh^{2}z_{0} + e^{+2
\sqrt{\epsilon}   t}}   , \nonumber\\
 z_{0} \neq 0   ,\qquad
t \rightarrow + \infty   , \qquad \phi - \phi _{0}  = { \alpha
\over 2 \sqrt{\epsilon}}   ( + \infty   )  . %\label{3.12b}
\end{gather*}

Before going farther,
let us note that from equation (\ref{1.7a}) it follows the conservation of squared velocity (or the energy)
in the presence of a magnetic
f\/ield
\begin{gather*}
\epsilon = \cosh^{2}z  \left [  \left({d r \over dt}\right)^{2} +
\sinh^{2}r  \left({d \phi \over dt}\right)^{2}   \right ] + \left({d z \over
dt}\right)^{2}
  .
 %\label{4.7}
 \end{gather*}
For trajectories with constant  $r= r_{0}$, the energy
looks  simpler (which coincides with  (\ref{3.3}))
\begin{gather*}
\epsilon  =    \sinh^{2}r_{0}
 {  B^{2} \over  \cosh^{2} r_{0} }   {1 \over \cosh^{2}   z }
  +   \left({d z \over dz}\right)^{2} =
  {A \over \cosh^{2}   z }    +   \left({d z \over dz}\right)^{2}   .
%\label{4.8c}
\end{gather*}

The above  elementary treatment seems not to be completely
satisfactory because we cannot be sure that all possible  motions
of the particle in a magnetic  f\/ield in the Lobachevsky space
have  been found. So  we turn to the  Lagrange formalism.

\section[Particle in a magnetic field and Lagrange formalism
in  $H_{3}$]{Particle in a magnetic f\/ield and Lagrange formalism
in  $\boldsymbol{H_{3}}$}

Now, let us consider the problem using the Lagrange formalism
\cite{Landau-Lifshitz-2}
\begin{gather*}
L = {1 \over 2}     \big(- g_{ik} V^{i}V^{k} \big)  -      g_{ik}
A^{i}V^{k} \nonumber
\\
\phantom{L}={1 \over 2}   \big(   \cosh^{2}z   V^{r}  V^{r}  +
\cosh^{2}z  \sinh^{2}r  V^{\phi}  V^{\phi}
 +   V^{z} V^{z}   \big) +  B   ( \cosh  r -1 ) V^{\phi}   .
%\label{5.1}
\end{gather*}
Euler--Lagrange equations read
\begin{gather}
{d \over dt}   \cosh^{2} z V^{r}  =   \cosh^{2} z
\sinh  r  \cosh  r  V^{\phi} V^{\phi} +  B
\sinh  r   V^{\phi}   , \nonumber
\\
{d \over dt}  \big[  \cosh^{2}z  \sinh^{2}r  V^{\phi}  +
 B   ( \cosh  r -1 )  \big]  = 0   , \nonumber\\
{d \over dt}   V^{z} = \cosh  z   \sinh  z    \big(
V^{r}  V^{r}  +  \sinh^{2}r  V^{\phi}  V^{\phi} \big)   ,
\label{5.2}
\end{gather}
or
\begin{gather*}
   {d V^{r}  \over d t   } +
2 \tanh z   V^{r}  V^{z} - \sinh  r   \cosh  r
  V^{\phi }  V^{\phi} =
  B   { \sinh  r    \over \cosh^{2} z }  V^{\phi}      ,
\nonumber
\\
{d  V^{\phi}  \over d t  } + 2   \coth  r   V^{\phi }
V^{r}  +   2   \tanh  z    V^{\phi } V^{z} =
 - B    {1   \over \cosh^{2} z   \sinh   r}    V^{r}     ,
\nonumber
\\
{d  V^{z}  \over d t  }  = \sinh  z  \cosh  z     \big(
V^{r}  V^{r} +  \sinh^{2} r   V^{\phi}  V^{\phi}   \big)     ,
%\label{5.3}
\end{gather*}

\noindent which coincide with equations~(\ref{2.12}). Second equation
in (\ref{5.2})  evidently determines a new conserved quantity
\begin{gather}
I =   \cosh^{2}z  \sinh^{2}r  V^{\phi}  +
 B    ( \cosh  r -1 ) = \mbox{const}   .
\label{5.4}
\end{gather}

We  will   obtain more from Lagrange formalism, if we  use  three
integrals of motion. Two of them are already  known
\begin{gather*}
I =   \cosh^{2}z  \sinh^{2}r  V^{\phi}  +
 B    ( \cosh  r -1 )     , \nonumber
 \\
 \epsilon = \cosh^{2}z   \big( V^{r}  V^{r}  +  \sinh^{2}r  V^{\phi}  V^{\phi}   \big)  +   V^{z} V^{z}  .
%\label{5.5}
\end{gather*}
Having remembered equation~(\ref{3.3}), it is easy to
guess the third
\begin{gather}
A = \cosh^{2} z   \left[    \epsilon - \left({d z \over dt}\right)^{2}    \right]
= \cosh^{4} z   \big(       V^{r}  V^{r}  +   \sinh^{2}r
V^{\phi}  V^{\phi}   \big)   . \label{5.6}
\end{gather}
To see that  $A$ indeed conserves  it suf\/f\/ices to rewrite
$A$ as
\begin{gather}
A = \big( \cosh^{2} z V^{r}\big)^{2} + {1 \over  \sinh^{2}r}  \big (
\cosh^{2} z   \sinh^{2} r  V^{\phi}\big)^{2} , \nonumber
\end{gather}
and takes into account the f\/irst and second equations in
(\ref{5.2})
\begin{gather}
{d \over dt}  \big(  \cosh^{2} z V^{r}  \big)  =   \cosh^{2}
z   \sinh  r  \cosh  r V^{\phi} V^{\phi} +  B
\sinh  r  V^{\phi}   ,
\nonumber
\\
{d \over dt} \big(  \cosh^{2}z  \sinh^{2}r V^{\phi}
)  = -    B   \sinh\; r V^{r}  = 0   , \nonumber
\end{gather}

\noindent then we arrive at
$d A / d t   = 0$.
With the help of three integrals of motion we can reduce the
problem in its most  general form (without any additional and
simplifying  assumptions) to calculating several integrals.
Indeed, from (\ref{5.4}) it follows
\begin{gather}
 {d \phi \over d t} =  {1 \over  \cosh^{2} z  }
 { I - B   (\cosh  r -1 ) \over   \sinh^{2} r }   .
\label{5.9}
\end{gather}
Substituting it into  (\ref{5.6}) we get
\begin{gather*}
A =  \cosh^{4} z      \left({d r \over dt}\right)^{2}   +
   { [  I - B   (\cosh  r -1)  ]^{2} \over
  \sinh^{2} r }     ,
\end{gather*}
therefore
\begin{gather}
{d r \over dt} = \pm   {1 \over \cosh^{2} z  }   \sqrt { A
- { [   I - B   ( \cosh  r -1)   ] ^{2} \over
  \sinh^{2} r }  }   .
 \label{5.10}
\end{gather}

\noindent In turn,  from (\ref{5.6}) it follows
\begin{gather}
{dz \over dt} =  \pm   {1 \over \cosh  z}    \sqrt{
\epsilon  \cosh^{2} z - A}   . \label{5.11}
\end{gather}
Dividing (\ref{5.10}) by  (\ref{5.11}), we  obtain
\begin{gather*}
{  \sinh   r   d r \over  \sqrt   { A   \sinh^{2} r  -
  (I - B   \cosh  r + B)^{2}  } } =  \pm  {1 \over \cosh   z   }
  { dz  \over \sqrt{ \epsilon  \cosh^{2} z - A}}      ,
%\label{5.12a}
\end{gather*}
which represents  trajectory equation in the form
 $d F(r,z)=0$.  In turn, dividing (\ref{5.10})  by (\ref{5.9}),
we get  trajectory  equation in the form   $dF(r,\phi)=0$
\begin{gather*}
{ [   I - B( \cosh  r -1)   ]   dr \over \sinh  r
\sqrt{ A  \sinh^{2}r - [   I - B( \cosh  r -1)  ]^{2}
}} = d \phi   . %\label{5.12b}
\end{gather*}

Thus, the solution of the problem~--   particle in a magnetic f\/ield on
the background of Lobachevsky space~-- reduces to   the following
integrals
\begin{gather}
 {d \phi \over d t} =  {1 \over  \cosh^{2} z  }  { I - B   (\cosh  r -1 ) \over   \sinh^{2} r }   ,
\label{5.13a}
\\
{d r \over dt} = \pm   {1 \over \cosh^{2} z  }   \sqrt { A
- { [   I - B   ( \cosh  r -1)   ] ^{2} \over
  \sinh^{2} r }  }  ,
\label{5.13b}
\\
{dz \over dt} =  \pm   {1 \over \cosh  z}    \sqrt{
\epsilon  \cosh^{2} z - A}   , \label{5.13c}
\\
{  \sinh   r   d r \over  \sqrt   { A   \sinh^{2} r  -
  (I - B  \cosh  r + B)^{2}  } } =  \pm  {1 \over \cosh   z   }
   { dz  \over \sqrt{ \epsilon \cosh^{2} z - A}}     ,
\label{5.13d}
\\
\pm   { [   I - B( \cosh  r -1)   ]   dr \over
\sinh  r  \sqrt{ A  \sinh^{2}r - [   I - B(
\cosh  r -1)  ]^{2} }} = d \phi   . \label{5.13e}
\end{gather}

The last f\/ive relations are valid for all possible solutions of
the problem under consideration.
In particular, let us show that the
  restriction  $r = r_{0} = \mbox{const}$ is compatible with  equations~(\ref{5.13a})--(\ref{5.13e}).
Indeed they give
\begin{gather}
 {d \phi \over d t} = {\alpha  \over  \cosh^{2} z  }
  , \qquad {dz \over dt} =  \pm   {1 \over \cosh  z}    \sqrt{
\epsilon  \cosh^{2} z - A}      , \nonumber
\\
 \alpha =   { I - B   (\cosh r_{0} -1 ) \over   \sinh^{2} r_{0} }    ,\qquad
 A =     { [  I - B   (\cosh  r_{0} -1)  ]^{2} \over
  \sinh^{2} r_{0} }     .
\label{5.14}
\end{gather}

It should be noted that in Section~\ref{section3} we had other
representations for  $\alpha$ and  $A$
\begin{gather}
\alpha = -   { B  \over \cosh  r_{0}}   , \qquad A =
\tanh^{2} r_{0}     B^{2}     . \label{5.15}
\end{gather}
However they are equivalent. Indeed, equating two
expressions for $\alpha $
\begin{gather}
{ I - B    (\cosh  r_{0} -1 ) \over   \sinh^{2} r_{0} }
= -   { B  \over \cosh  r_{0}}  \quad \Longrightarrow \quad I = B
{1 - \cosh  r_{0} \over  \cosh  r_{0}  }   ,
\label{5.16}
\end{gather}
and substituting this  into $A$ in  (\ref{5.14}) we get
\begin{gather*}
A =     { [  I - B   (\cosh  r_{0} -1)  ]^{2} \over
  \sinh^{2} r_{0} } \\
  \phantom{A} = { B^{2} \over  \sinh^{2} r_{0} }
  \left[  {(1 - \cosh  r_{0}) \over  \cosh  r_{0} } + (1 - \cosh  r_{0})  \right ]^{2} =
   \big(   \tanh^{2} r_{0}    B^{2} \big)   ,
\end{gather*}
which coincides with  (\ref{5.15}). It should be
specially noted that from (\ref{5.16}) we  must conclude that to
the case of the most simple  motion  $r_{0} =
\mbox{const}$, there corresponds the inequality
\begin{gather}
\cosh  r_{0} = {1 \over 1 + I / B} >1   ,\qquad  \mbox{that
is } \quad {B \over  I + B} > 1    . \label{5.17}
\end{gather}
From (\ref{5.17}) we conclude that the motion with
$r=r_{0}$ is  possible if
\begin{gather*}
B > 0   , \qquad  - B < I < 0;     \qquad
B<0 , \qquad  0 < I < - B   . %\label{I-values}
\end{gather*}

\section[Possible solutions in $H_{3}$, radial finite and infinite motions]{Possible solutions in $\boldsymbol{H_{3}}$, radial f\/inite and inf\/inite motions}\label{section6}

Now, with the use of general relationships   (\ref{5.13a})--(\ref{5.13e}), let us examine the general case without restriction  $r
= r_{0}$. Let us   $A  \neq \epsilon $.

First, as shown above equation  (\ref{5.13c})  results in
\begin{gather*}
{\rm I.}   \qquad
\epsilon > A  , \qquad  z \in ( - \infty, + \infty )    , \qquad
\sinh\; z (t) = \pm   \sqrt{ 1 - {A  \over \epsilon}  }
\sinh  \sqrt{\epsilon}   t  ; \nonumber
\\
{\rm II.}    \qquad
 A > \epsilon  , \qquad   \sinh^{2} z > {A  \over  \epsilon } -1    , \qquad
 \sinh\; z (t) = \pm
\sqrt{{ A  \over \epsilon }-1}     \cosh   \sqrt{\epsilon}
  t     . %\label{6.1b}
\end{gather*}
Equation  (\ref{5.13b}) can be rewritten as
\begin{gather}
\pm   \int   { d     \cosh   r   \over  \sqrt   { A
(\cosh^{2} r -1 )  -
  (I +B - B   \cosh  r )^{2}  } } =
  \int {d t \over \cosh^{2} z  (t) }     .
 \label{6.3}
 \end{gather}
Integral in the right-hand side is known~-- see
(\ref{3.10a})
\begin{gather*}
{\rm I.}   \qquad
\epsilon > A  , \qquad R  =  \int {d t \over \cosh^{2} z }
 = {  1  \over \sqrt{ A} } \, \mbox{arccoth}\left ( \sqrt{{A \over \epsilon}}
\tanh  \sqrt{\epsilon}  t \right)  , \nonumber
\\
{\rm  II.}    \qquad
\epsilon < A   , \qquad R   =  \int {d t \over \cosh^{2} z }
 = {  1  \over \sqrt{ A} } \,  \mbox{arccoth}\left( \sqrt{{\epsilon \over A}}
\tanh   \sqrt{\epsilon}  t \right)    . %\label{6.4}
\end{gather*}
For the integral in the left-hand side we   get
\begin{gather}
L = \int  {  dx  \over  \sqrt   { a x^{2} + b x + c } } =
 \int
{dx \over \sqrt{ a  (x +{b\over 2a})^{2} + {b^{2}-4ac \over
-4a}}} , \qquad x = \cosh r , \nonumber
\\
 a = A -  B^{2} , \qquad
b = 2B (I + B)  , \qquad c = -A - (I+B)^{2} < 0   .
\label{6.5a}
\end{gather}

It should be noted the identity
\begin{gather*}
a+b+c = - I^{2}   . %\label{6.5a'}
\end{gather*}
Depending on the values of $a$, $b$, $c$ there may be realized
solutions of dif\/ferent types.

\subsection{Finite radial motions}
Let it be $a < 0$, $  B^{2} - A > 0 $.
The roots $x_{1}$ and $ x_{2}$  are
\begin{gather*}
x_{1} = {b - \sqrt{b^{2}-4ac} \over -2a}   , \qquad x_{2} = {b +
\sqrt{b^{2}-4ac} \over -2a}  , \\
b^{2} - 4ac = 4A   \big[   (I+B)^{2} - (B^{2} - A)   \big]   ,
%\label{rout-1a}
\end{gather*}
and  inequality $ b^{2}-4ac > 0$ reduces to the
following restriction
$
(I +B)^{2} -( B^{2} - A ) > 0 $. In general, at $a<0$  we might expect  several
possibilities

\medskip

{\centering \begin{minipage}{75mm}
%\unitlength=0.5 mm
%\begin{picture}(160,100)(-40,-60)
%\special{em:linewidth 0.4pt} \linethickness{0.4pt}
%
%
%\put(-20,0){\vector(+1,0){80}}     \put(+60,-5){$x=\cosh  r
%$} \put(0,-30,0){\vector(0,+1){60}}   \put(+5,+30){$y(x) = ax^{2}
%+ bx + c >0  $}
%
%
%\put(35,+19){\circle*{1}} \put(30,+16){\circle*{1}}
%\put(40,+16){\circle*{1}} \put(28,+13){\circle*{1}}
%\put(42,+13){\circle*{1}} \put(26,+10){\circle*{1}}
%\put(44,+10){\circle*{1}} \put(22,+6){\circle*{1}}
%\put(48,+6){\circle*{1}} \put(21,+2){\circle*{1}}
%\put(49,+2){\circle*{1}} \put(18,-3){\circle*{1}}
%\put(52,-3){\circle*{1}} \put(17,-7){\circle*{1}}
%\put(53,-7){\circle*{1}} \put(16,-10){\circle*{1}}
%\put(54,-10){\circle*{1}}
%
%
%\put(10,-0.5){\circle*{2}}  \put(8,-8){$1$}
%\put(20,-0.5){\circle*{2}}  \put(18,+5){$x_{1}$}
%\put(50,-0.5){\circle*{2}}  \put(48,+5){$x_{2}$}
%
%\put(20,-0.3){\line(+1,0){30} } \put(20,-0.5){\line(+1,0){30} }
%\put(20,-0.8){\line(+1,0){30} }
%
%\put(+60,+15){$ 1 < x_{1} \leq  x  \leq x_{2}$}
%
%\end{picture}
\centerline{\includegraphics{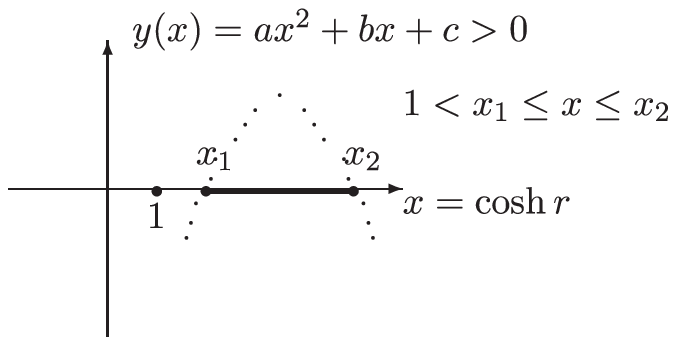}}
\centerline{{\bf Fig.~1a.} Finite  motion.}
\end{minipage}
\qquad
\begin{minipage}{75mm}
%\unitlength=0.5 mm
%\begin{picture}(160,100)(-40,-60)
%\special{em:linewidth 0.4pt} \linethickness{0.4pt}
%
%\put(25,+19){\circle*{1}} \put(20,+16){\circle*{1}}
%\put(30,+16){\circle*{1}} \put(18,+13){\circle*{1}}
%\put(32,+13){\circle*{1}} \put(16,+10){\circle*{1}}
%\put(34,+10){\circle*{1}} \put(12,+6){\circle*{1}}
%\put(38,+6){\circle*{1}} \put(11,+2){\circle*{1}}
%\put(39,+2){\circle*{1}} \put(8,-3){\circle*{1}}
%\put(42,-3){\circle*{1}} \put(7,-7){\circle*{1}}
%\put(43,-7){\circle*{1}} \put(6,-10){\circle*{1}}
%\put(44,-10){\circle*{1}}
%\put(-20,0){\vector(+1,0){80}}     \put(+60,-5){$x=\cosh r
%$} \put(0,-30,0){\vector(0,+1){60}}   \put(+5,+30){$y(x)= ax^{2} +
%bx + c   >0$} \put(10,-0.5){\circle*{2}}  \put(8,-8){$1$}
%\put(10,-0.5){\circle*{2}}  \put(8,+5){$x_{1}$}
%\put(40,-0.5){\circle*{2}}  \put(38,+5){$x_{2}$}
%
%\put(10,-0.3){\line(+1,0){30} } \put(10,-0.5){\line(+1,0){30} }
%\put(10,-0.7){\line(+1,0){30} }
%
%\put(+60,+15){$ 1 =x_{1} \leq  x  \leq x_{2}  $ }
%
%\end{picture}
\centerline{\includegraphics{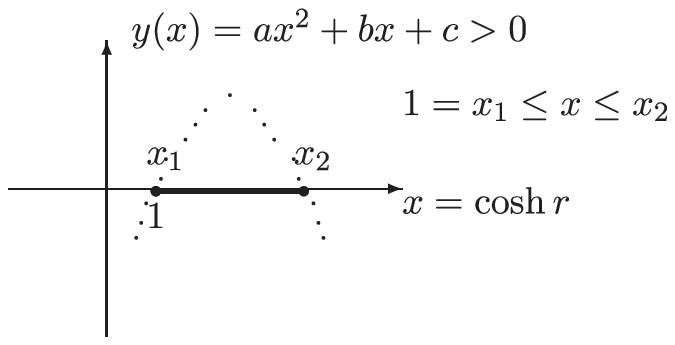}}
\centerline{{\bf Fig.~1b.} Finite motion.}
\end{minipage}
}

\bigskip

{\centering\begin{minipage}{75mm}
%\begin{picture}(160,100)(-40,-60)
%\unitlength=0.5 mm
%\special{em:linewidth 0.4pt} \linethickness{0.4pt}
%
%
%\put(20,+19){\circle*{1}} \put(15,+16){\circle*{1}}
%\put(25,+16){\circle*{1}} \put(13,+13){\circle*{1}}
%\put(27,+13){\circle*{1}} \put(11,+10){\circle*{1}}
%\put(29,+10){\circle*{1}} \put(7,+6){\circle*{1}}
%\put(33,+6){\circle*{1}} \put(6,+2){\circle*{1}}
%\put(34,+2){\circle*{1}} \put(3,-3){\circle*{1}}
%\put(37,-3){\circle*{1}} \put(2,-7){\circle*{1}}
%\put(38,-7){\circle*{1}} \put(1,-10){\circle*{1}}
%\put(39,-10){\circle*{1}}
%
%
%\put(-20,0){\vector(+1,0){80}}     \put(+60,-5){$x=\cosh r
%$} \put(0,-30,0){\vector(0,+1){60}}   \put(+5,+30){$y(x)=  ax^{2}
%+ bx + c  > 0   $} \put(10,-0.5){\circle*{2}}  \put(8,-8){$1$}
%\put(5,-0.5){\circle*{2}}  \put(3,+5){$x_{1}$}
%\put(35,-0.5){\circle*{2}}  \put(33,+5){$x_{2}$}
%
%\put(10,-0.3){\line(+1,0){25} } \put(10,-0.5){\line(+1,0){25} }
%\put(10,-0.7){\line(+1,0){25} }
%
%\put(+60,+15){$ x_{1} < 1  \leq  x  \leq x_{2}$ }
%
%\end{picture}
\centerline{\includegraphics{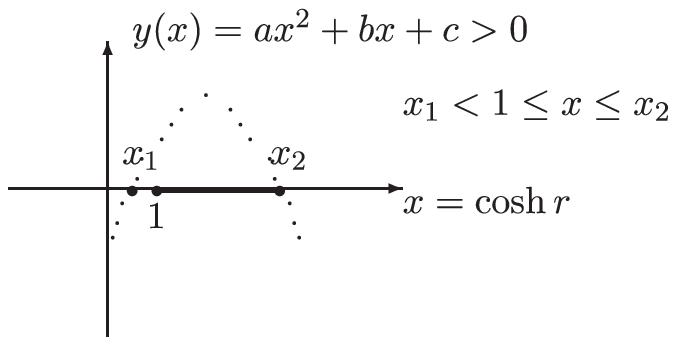}}

\centerline{{\bf Fig.~1c.} Finite motion.}
\end{minipage}
\qquad
\begin{minipage}{75mm}
%\vspace{+4mm} \unitlength=0.5 mm
%\begin{picture}(160,100)(-40,-60)
%\special{em:linewidth 0.4pt} \linethickness{0.4pt}
%
%
%
%\put(-5,+19){\circle*{1}} \put(-10,+16){\circle*{1}}
%\put(0,+16){\circle*{1}} \put(-12,+13){\circle*{1}}
%\put(2,+13){\circle*{1}} \put(-14,+10){\circle*{1}}
%\put(4,+10){\circle*{1}} \put(-18,+6){\circle*{1}}
%\put(8,+6){\circle*{1}} \put(-19,+2){\circle*{1}}
%\put(9,+2){\circle*{1}} \put(-22,-3){\circle*{1}}
%\put(12,-3){\circle*{1}} \put(-23,-7){\circle*{1}}
%\put(13,-7){\circle*{1}} \put(-24,-10){\circle*{1}}
%\put(14,-10){\circle*{1}}
%
%
%
%\put(-20,0){\vector(+1,0){80}}     \put(+60,-5){$x=\cosh  r
%$} \put(0,-30,0){\vector(0,+1){60}}   \put(+5,+30){$y(x)=ax^{2} +
%bx + c  > 0 $} \put(10,-0.5){\circle*{2}}  \put(8,-8){$1$}
%\put(-20,-0.5){\circle*{2}}  \put(-22,+5){$x_{1}$}
%\put(+10,-0.5){\circle*{2}}  \put(+8,+5){$x_{2}=1$}
%
%\put(+60,+15){$x =1$ }
%
%\end{picture}
\centerline{\includegraphics{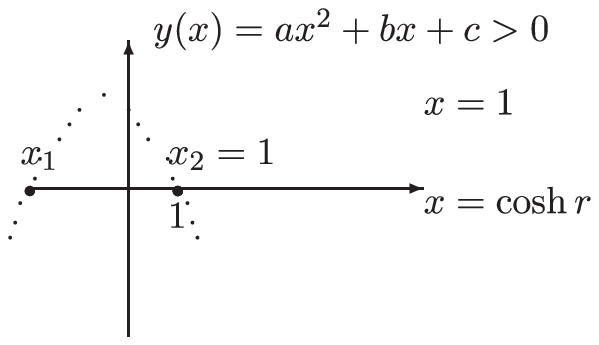}}
\centerline{{\bf Fig.~1d.} Very special state  $r=0$.}
\end{minipage}
}

\bigskip

\centerline{\begin{minipage}{75mm}
%\vspace{+4mm} \unitlength=0.5 mm
%\begin{picture}(160,100)(-40,-60)
%\special{em:linewidth 0.4pt} \linethickness{0.4pt}
%
%
%\put(-20,0){\vector(+1,0){80}}     \put(+60,-5){$x=\cosh  r
%$} \put(0,-30,0){\vector(0,+1){60}}   \put(+5,+30){$y(x)= ax^{2} +
%bx + c  > 0$}
%
%
%\put(-10,+19){\circle*{1}} \put(-15,+16){\circle*{1}}
%\put(-5,+16){\circle*{1}} \put(-17,+13){\circle*{1}}
%\put(-3,+13){\circle*{1}} \put(-19,+10){\circle*{1}}
%\put(-1,+10){\circle*{1}} \put(-23,+6){\circle*{1}}
%\put(3,+6){\circle*{1}} \put(-24,+2){\circle*{1}}
%\put(4,+2){\circle*{1}} \put(-27,-3){\circle*{1}}
%\put(7,-3){\circle*{1}} \put(-28,-7){\circle*{1}}
%\put(8,-7){\circle*{1}} \put(-29,-10){\circle*{1}}
%\put(9,-10){\circle*{1}}
%
%
%\put(10,-0.5){\circle*{2}}  \put(8,-8){$1$}
%\put(-25,-0.5){\circle*{2}}  \put(-28,+5){$x_{1}$}
%\put(+5,-0.5){\circle*{2}}  \put(+2,+5){$x_{2}$}
%\end{picture}
\centerline{\includegraphics{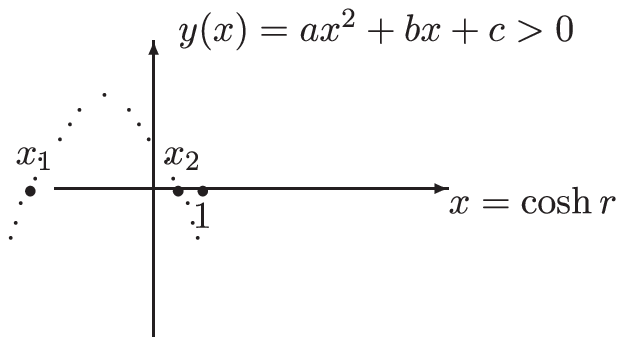}}
\centerline{{\bf Fig.~1e.}  No physical solution.}
\end{minipage}}

\bigskip

Physically  interesting   cases, Figs.~1a--1c, can be
characterized additionally by constrains on $a$, $b$, $c$
 \begin{alignat*}{3}
& \mbox{Fig. 1a}   \qquad &&   b +2a  > 0    ,\qquad a + b +
c < 0 \qquad  \Longrightarrow  & \nonumber
\\
&&&  BI+A > 0   , \qquad -I^{2}   < 0
\qquad  (\mbox{true inequality}) ; \nonumber &
\\
& \mbox{Fig. 1b}   \qquad &&  b +2a > 0   ,\qquad a + b +
c = 0 \qquad  \Longrightarrow \qquad A > 0   , \qquad I   = 0
\nonumber&
\\
& \mbox{Fig. 1c}   \qquad &&
 a + b + c > 0 \qquad   \Longrightarrow \qquad
-I^{2}   > 0 \qquad \mbox{(false statement)}. &
 %\label{rout-1b'}
\end{alignat*}
Therefore, only  Figs.~1a and~1b correspond to the
physically possible solutions (with two  turning points in radial
variable)
\begin{gather}
\mbox{Fig. 1a},  \qquad  B^{2} - A  > 0   , \qquad
 (I +B)^{2} -( B^{2} - A ) > 0    , \qquad BI+A > 0  , \qquad I \neq 0;
 \nonumber
 \\
 \mbox{Fig. 1b}, \qquad  B^{2} - A  > 0   , \qquad
  A > 0   , \qquad I = 0.
\label{rout-1b''}
\end{gather}

Now let us turn to the integral  (\ref{6.5a}) when
(\ref{rout-1b''}) holds
\begin{gather*}
L =  \int {dx \over \sqrt{ a   (x +{b\over 2a})^{2} + {b^{2}-4ac
\over -4a}}}=
 {1 \over \sqrt{-a}} \arcsin { -2a x - b  \over
\sqrt{b^{2} - 4ac}}  , \qquad x_{1} < x < x_{2}   . %\label{6.10}
\end{gather*}
Therefore, equation~(\ref{6.3})  gives (let $\epsilon > A$)
\begin{gather*}
 {1 \over \sqrt{-a}} \, \mbox{arcsin}\, { -2a \; \cosh\;r  - b  \over
\sqrt{b^{2} - 4ac}} =\pm
 {  1  \over \sqrt{ A} } \, \mbox{arccoth}  \left( \sqrt{{A \over \epsilon}}
\tanh   \sqrt{\epsilon}  t \right) ,
\nonumber
\end{gather*}
or
\begin{gather}
 { -2a   \cosh r  - b  \over
\sqrt{b^{2} - 4ac}} = \pm
 \sin    \left [ {  \sqrt{-a}   \over \sqrt{ A} } \,  \mbox{arccoth}\, \left( \sqrt{{A \over \epsilon}}
\tanh   \sqrt{\epsilon}  t \right)      \right ]
 .
\label{6.11a}
\end{gather}
As expected, the variable $\cosh   r$ runs within
a f\/inite segment
\begin{gather*}
x_{1} ={b - \sqrt{b^{2}- 4ac } \over -2a } \leq \cosh   r
\leq  {b + \sqrt{b^{2}- 4ac } \over -2a } = x_{2}   ,
\end{gather*}
or (see Fig.~1a)
\begin{gather*}
B^{2}-A > 0 , \qquad (I+B)^{2} -( B^{2} -A ) > 0   , \qquad BI
+A > 0   , \nonumber
\\
{2B(I+B) - \sqrt{4A   [   (I+B)^{2} - (B^{2}-A)   ]} \over
2(B^{2}-A) } \leq \cosh   r    , \nonumber
\\
\cosh   r \leq {2B(I+B) + \sqrt{4A  [   (I+B)^{2} - (B^{2}-
A)  ]} \over 2(B^{2}-A) }    , %\label{6.11b}
\end{gather*}
which at $I=0$  takes a more simple form (see Fig.~1b)
\begin{gather*}
A< B^{2}   , \qquad A > 0 , \qquad 1 \leq  \cosh   r   \leq
{B^{2} + A \over B^{2} - A }. %\label{6.11c}
\end{gather*}

It should be noted that when
\begin{gather*}
b^{2}- 4ac=0 \qquad \mbox{or equivalently} \qquad B^{2} - A = (I +
B)^{2}  \label{6.11c}
\end{gather*}
according to  (\ref{6.11a})  the motion within the segment  $[ x_{1}, x_{2} ]$
 reduces to the
  motion with a f\/ixed  value $r_{0}$
\begin{gather*}
-2a   \cosh r_{0}  - b = 0 \qquad  \Longrightarrow \qquad
    \cosh r_{0}  = {b \over  -2 a}=
{2B (I + B) \over 2 (B^{2} - A) } = {B \over  I + B } ,
%\label{6.11d}
\end{gather*}
which coincides with the expression for $\cosh r_{0}$
given by (\ref{5.16}).

\subsection[Infinite radial motions ($a>0$)]{Inf\/inite radial motions ($\boldsymbol{a>0}$)}

One special case arises when $a=A-B^{2}=0$,  indeed then we have
\begin{gather}
L = \int {dx \over \sqrt{bx + c } } = {2 \over b}   \sqrt{b x +
c}    , \qquad b > 0   , \qquad x = \cosh  r \geq - {c
\over b}    , \nonumber
\\
-{c \over b} = {B^{2} + (I+B)^{2} \over 2B(I+B) } \geq +1 \quad
\Longrightarrow \quad B(I+B)> 0   , \label{addition-1}
\end{gather}
which  corresponds to  an inf\/inite motion in radial
variable (let $\epsilon > A$)
\begin{gather}
{2 \over \sqrt{b}}   \sqrt{ \cosh   r  + {c \over b}} = \pm
  {  1  \over \sqrt{ B^{2}} } \, \mbox{arccoth}\left(  \sqrt{{B^{2} \over \epsilon}}
\tanh   \sqrt{\epsilon}  t  \right)   . \label{addition-2}
\end{gather}

Now let us  examine  other possibilities related  to  inequality
$a>0$. When
$a = A -B^{2} >0 $,
  the roots $x_{1}$, $x_{2}$  are def\/ined by
\begin{gather*}
x_{1} = {- b - \sqrt{b^{2}-4ac} \over 2a}   , \qquad x_{2} = {- b
+ \sqrt{b^{2}-4ac} \over 2a}  , %\label{rout-2a}
\end{gather*}
and inequality $ b^{2}-4ac > 0$ reduces to
\begin{gather*}
(I +B)^{2} + (A - B^{2} ) >  0   . %\label{rout-2b}
\end{gather*}

In general, at the restrictions  $a > 0$  we might expect the
following cases

%\vspace{4mm} \unitlength=0.45 mm
%\begin{picture}(160,100)(-40,-60)
%\special{em:linewidth 0.4pt} \linethickness{0.4pt}
%
%
%\put(-20,0){\vector(+1,0){80}}     \put(+60,-7){$x=\cosh r
%$} \put(0,-30,0){\vector(0,+1){60}}   \put(+5,+30){$y(x) = ax^{2}
%+ bx + c > 0 $}
%
%
%\put(-10,-19){\circle*{1}} \put(-15,-16){\circle*{1}}
%\put(-5,-16){\circle*{1}} \put(-17,-13){\circle*{1}}
%\put(-3,-13){\circle*{1}} \put(-19,-10){\circle*{1}}
%\put(-1,-10){\circle*{1}} \put(-23,-6){\circle*{1}}
%\put(3,-6){\circle*{1}} \put(-24,-2){\circle*{1}}
%\put(4,-2){\circle*{1}} \put(-27,+3){\circle*{1}}
%\put(7,+3){\circle*{1}} \put(-28,+7){\circle*{1}}
%\put(8,+7){\circle*{1}} \put(-29,+10){\circle*{1}}
%\put(9,+10){\circle*{1}}
%
%
%\put(10,-0.5){\circle*{2}}  \put(8,-8){$1$}
%\put(-25,-0.5){\circle*{2}}  \put(-28,+5){$x_{1}$}
%\put(+5,-0.5){\circle*{2}}  \put(+2,+5){$x_{2}$}
%
%\put(10,-0.5){\line(+1,0){50} } \put(10,-0.8){\line(+1,0){50} }
%\put(10,-0.10){\line(+1,0){50} }
%
%\put(+60,+15){$  x \geq +1$ }
%
%\end{picture}
\bigskip

{\centering
\begin{minipage}{75mm}
\centerline{\includegraphics{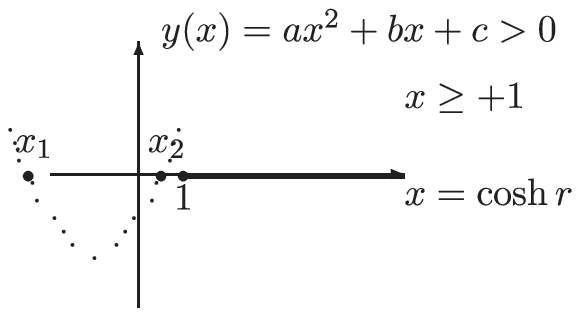}}

\centerline{{\bf Fig.~2a.} Inf\/inite motion.}

\medskip

\centerline{$   b+2a  > 0  ,  \quad a +b + c > 0$}
\centerline{$\Longrightarrow\quad
   BI + A > 0 , \quad  -I^{2} > 0 \quad \mbox{Impossible}$}
\end{minipage}\qquad
%\unitlength=0.45 mm
%\begin{picture}(160,100)(-40,-60)
%\special{em:linewidth 0.4pt} \linethickness{0.4pt}
%
%
%\put(-20,0){\vector(+1,0){80}}     \put(+60,-7){$x=\cosh  r
%$} \put(0,-30,0){\vector(0,+1){60}}   \put(+5,+30){$y(x)= ax^{2} +
%bx + c  > 0 $}
%
%
%\put(-5,-19){\circle*{1}} \put(-10,-16){\circle*{1}}
%\put(0,-16){\circle*{1}} \put(-12,-13){\circle*{1}}
%\put(+2,-13){\circle*{1}} \put(-14,-10){\circle*{1}}
%\put(+4,-10){\circle*{1}} \put(-18,-6){\circle*{1}}
%\put(8,-6){\circle*{1}} \put(-19,-2){\circle*{1}}
%\put(9,-2){\circle*{1}} \put(-22,+3){\circle*{1}}
%\put(12,+3){\circle*{1}} \put(-23,+7){\circle*{1}}
%\put(13,+7){\circle*{1}} \put(-24,+10){\circle*{1}}
%\put(14,+10){\circle*{1}}
%
%
%\put(10,-0.5){\circle*{2}}  \put(8,-8){$1$}
%\put(-25,-0.5){\circle*{2}}  \put(-28,+5){$x_{1}$}
%\put(+10,-0.5){\circle*{2}}  \put(+5,+5){$x_{2}$}
%
%\put(10,-0.5){\line(+1,0){50} } \put(10,-0.8){\line(+1,0){50} }
%\put(10,-0.10){\line(+1,0){50} }
%
%\put(+60,+15){$  x \geq x_{2} = +1 $  }
%
%\end{picture}
\begin{minipage}{75mm}
\centerline{\includegraphics{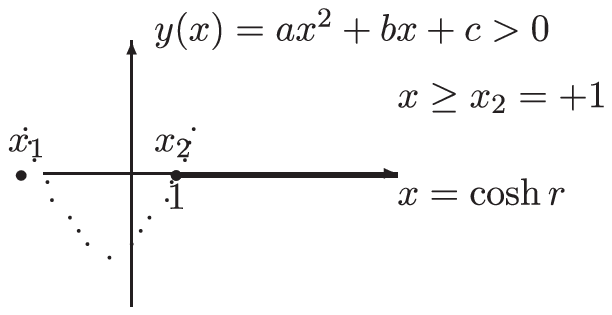}}

\centerline{{\bf Fig.~2b.} Inf\/inite motion.}

\medskip

\centerline{$b  + 2a > 0, \quad   a + b + c = 0$}
 \centerline{$\Longrightarrow \quad A > 0, \quad I = 0 \quad \mbox{Possible}$}
\end{minipage}
}

%\vspace{4mm}
%\unitlength=0.45 mm
%\begin{picture}(160,100)(-40,-60)
%\special{em:linewidth 0.4pt} \linethickness{0.4pt}
%
%
%\put(-20,0){\vector(+1,0){90}}     \put(+60,-7){$x=\cosh r
%$} \put(0,-30,0){\vector(0,+1){60}}   \put(+5,+30){$y(x)= ax^{2} +
%bx + c  $}
%
%
%\put(20,-19){\circle*{1}} \put(15,-16){\circle*{1}}
%\put(25,-16){\circle*{1}} \put(13,-13){\circle*{1}}
%\put(27,-13){\circle*{1}} \put(11,-10){\circle*{1}}
%\put(29,-10){\circle*{1}} \put(7,-6){\circle*{1}}
%\put(33,-6){\circle*{1}} \put(6,-2){\circle*{1}}
%\put(34,-2){\circle*{1}} \put(3,+3){\circle*{1}}
%\put(37,+3){\circle*{1}} \put(2,+7){\circle*{1}}
%\put(38,+7){\circle*{1}} \put(1,+10){\circle*{1}}
%\put(39,+10){\circle*{1}}
%
%
%\put(10,-0.5){\circle*{2}}  \put(8,-8){$1$}
%\put(5,-0.5){\circle*{2}}  \put(2,+5){$x_{1}$}
%\put(35,-0.5){\circle*{2}}  \put(32,+5){$x_{2}$}
%
%\put(35,-0.2){\line(+1,0){30} } \put(35,-0.4){\line(+1,0){30} }
%\put(35,-0.7){\line(+1,0){30} }
%
%\put(+60,+15){$ x >  x_{2} > +1$}
%
%\end{picture}

\bigskip

{\centering
\begin{minipage}{75mm}
\centerline{\includegraphics{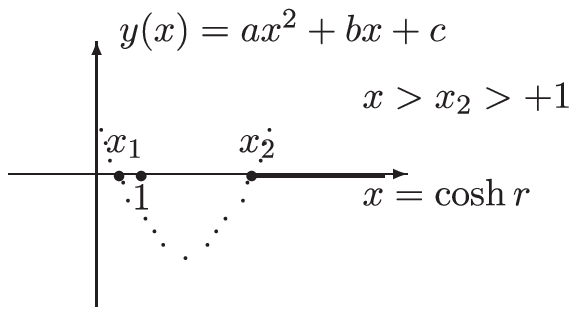}}

\centerline{{\bf Fig.~2c.} Inf\/inite motion.}

\medskip

\centerline{$a + b + c < 0$}

\centerline{$\Longrightarrow \quad
  -I^{2} < 0 \quad \mbox{Possible}$}
\end{minipage}
\qquad
%\vspace{4mm}
%\unitlength=0.45 mm
%\begin{picture}(160,100)(-40,-60)
%\special{em:linewidth 0.4pt} \linethickness{0.4pt}
%
%
%\put(-20,0){\vector(+1,0){88}}     \put(+60,-7){$x=\cosh  r
%$} \put(0,-30,0){\vector(0,+1){60}}   \put(+5,+30){$y(x)=  ax^{2}
%+ bx + c  > 0 $}
%
%
%\put(25,-19){\circle*{1}} \put(20,-16){\circle*{1}}
%\put(30,-16){\circle*{1}} \put(18,-13){\circle*{1}}
%\put(32,-13){\circle*{1}} \put(16,-10){\circle*{1}}
%\put(34,-10){\circle*{1}} \put(12,-6){\circle*{1}}
%\put(38,-6){\circle*{1}} \put(11,-2){\circle*{1}}
%\put(39,-2){\circle*{1}} \put(8,+3){\circle*{1}}
%\put(42,+3){\circle*{1}} \put(7,+7){\circle*{1}}
%\put(43,+7){\circle*{1}} \put(6,+10){\circle*{1}}
%\put(44,+10){\circle*{1}}
%
%
%\put(10,-0.5){\circle*{2}}  \put(8,-8){$1$}
%\put(2,+5){$x_{1}=1$} \put(40,-0.5){\circle*{2}}
%\put(37,+5){$x_{2}$}
%
%\put(40,-0.3){\line(+1,0){24} } \put(40,-0.6){\line(+1,0){24} }
%\put(40,-0.9){\line(+1,0){24} }
%
%\put(+60,+15){$ x \geq x_{2}$  }
%
%\end{picture}
\begin{minipage}{75mm}
\centerline{\includegraphics{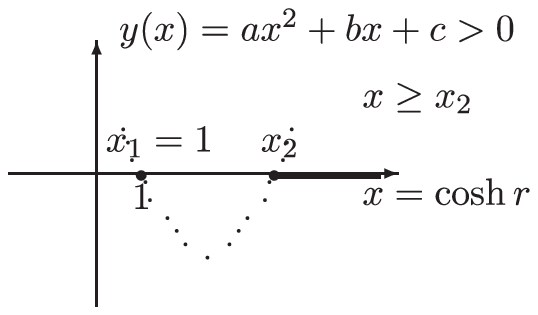}}

\centerline{{\bf Fig.~2d.} Inf\/inite motion.}

\medskip

\centerline{$2a + b < 0   , \quad a+b +c = 0$}

\centerline{$\Longrightarrow \quad
  A < 0 , \quad  I = 0 \quad \mbox{Impossible}$}
\end{minipage}}

\bigskip

\centerline{\includegraphics{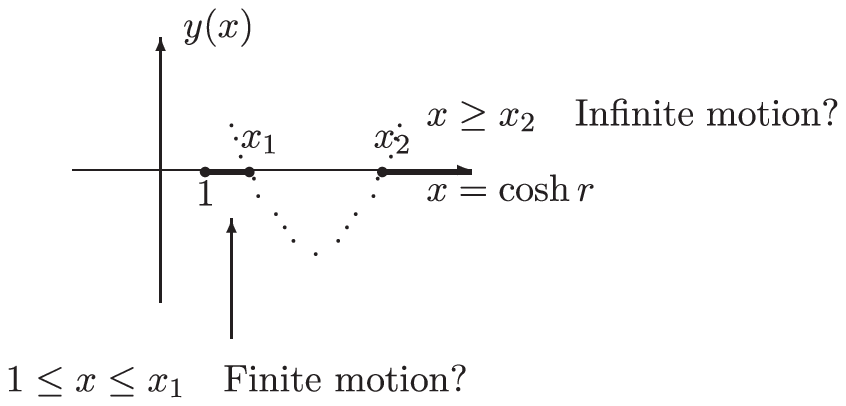}}

\centerline{{\bf Fig.~2e.} Nonphysical case.}

\medskip

% \unitlength=0.45 mm
%\begin{picture}(160,90)(-40,-60)
%\special{em:linewidth 0.4pt} \linethickness{0.4pt}
%
%
%\put(-20,0){\vector(+1,0){90}}     \put(+60,-7){$x=\cosh  r
%$} \put(0,-30,0){\vector(0,+1){60}}   \put(+5,+30){$y(x)$}
%
%
%\put(35,-19){\circle*{1}} \put(30,-16){\circle*{1}}
%\put(40,-16){\circle*{1}} \put(28,-13){\circle*{1}}
%\put(42,-13){\circle*{1}} \put(26,-10){\circle*{1}}
%\put(44,-10){\circle*{1}} \put(22,-6){\circle*{1}}
%\put(48,-6){\circle*{1}} \put(21,-2){\circle*{1}}
%\put(49,-2){\circle*{1}} \put(18,+3){\circle*{1}}
%\put(52,+3){\circle*{1}} \put(17,+7){\circle*{1}}
%\put(53,+7){\circle*{1}} \put(16,+10){\circle*{1}}
%\put(54,+10){\circle*{1}}
%
%
%\put(10,-0.5){\circle*{2}}  \put(8,-8){$1$}
%\put(20,-0.5){\circle*{2}}  \put(18,+5){$x_{1}$}
%\put(50,-0.5){\circle*{2}}  \put(48,+5){$x_{2}$}
%
%\put(10,-0.3){\line(+1,0){10} } \put(10,-0.6){\line(+1,0){10} }
%\put(10,-0.8){\line(+1,0){10} }
%
%
%\put(50,-0.3){\line(+1,0){20} } \put(50,-0.6){\line(+1,0){20} }
%\put(50,-0.8){\line(+1,0){20} }
%
%\put(-35,-50){$ 1  \leq  x  \leq x_{1} \ \  $   \mbox{Finite motion? }} \put(+16,-38){\vector(0,+1){27}}
%
%\put(+60,+10){$ x \geq x_{2} \ \ $    \mbox{Inf\/inite motion?
%} }
%
%\end{picture}

As concerns  Fig.~2e, we should  examine  the following
inequality{\samepage
\begin{gather*}
x_{1} > 1 \qquad  \Longrightarrow \qquad   -b >  \sqrt{b^{2} -
4ac} + 2a \qquad \Longrightarrow \qquad  (c-a ) >  +
 \sqrt{b^{2} - 4ac}  ,
\end{gather*}
which is impossible.}

 Therefore, only   Figs.~2b and~2c correspond to physically possible solutions
  (motions inf\/inite in radial variable $r$).
Integrating  (\ref{6.5a})  reduces to the elementary calculation
\begin{gather*}
L =
 \int
{dx \over \sqrt{ a   (x +{b\over 2a})^{2} + {b^{2}-4ac \over
-4a}}}=
 {1 \over \sqrt{a}} \, \mbox{arccosh}\, {2ax + b \over  \sqrt{b^{2} - 4ac}}  .
\end{gather*}
Therefore, equation  (\ref{6.3})  gives (with $\epsilon > A$)
\begin{gather*}
 {1 \over \sqrt{a}} \, \mbox{arccosh}\, { 2a   \cosh r  + b  \over
\sqrt{b^{2} - 4ac}} =\pm
 {  1  \over \sqrt{ A} } \,  \mbox{arccoth}\left( \sqrt{{A \over \epsilon}}
\tanh   \sqrt{\epsilon}  t \right)
 ,
%\label{6.11Ac}
\end{gather*}
or
\begin{gather}
{ 2a   \cosh r  + b  \over \sqrt{b^{2} - 4ac}} =
\cosh \left (
 {  \sqrt{a}  \over \sqrt{ A} } \,  \mbox{arccoth}\left ( \sqrt{{A \over \epsilon}}
\tanh   \sqrt{\epsilon}  t \right)   \right )
   ,
\label{6.11Ad}
\end{gather}
Evidently, equation (\ref{6.11Ad}) leads to (see Fig.~2c)
\begin{gather*}
\cosh  r > { - b + \sqrt{b^{2} -4ac} \over 2a }   , \qquad b
+ 2a  > 0   ,
\end{gather*}
or
\begin{gather*}
\cosh  r >  {-2B(I+B) + \sqrt{4A[ (I+B)^{2} + (A-B^{2})]}
\over 2 (A - B^{2} )}   , \qquad
 A+I B > 0  ,  \qquad B^{2} < A  ;
 %\label{d}
 \end{gather*}
from whence at $I=0$ it follows (see Fig.~2b)
$
\cosh  r >  +1$, $A > 0$.

\section[Trajectory equation in the form $F(r,z)=0$, model $H_{3}$]{Trajectory equation in the form $\boldsymbol{F(r,z)=0}$, model $\boldsymbol{H_{3}}$}\label{section7}

Now, let us consider the trajectory equation  $F(r,z)=0$ according
to (\ref{5.13d}).
\begin{gather*}
\int   {  \sinh   r   d r \over  \sqrt   { A
\sinh^{2} r  -
  (I - B   \cosh  r + B)^{2}  } } =  \pm  \int   {1 \over \cosh   z   }
   { dz  \over \sqrt{ \epsilon  \cosh^{2} z - A}}  .
\end{gather*}

Its right-hand side gives
\begin{gather*}
{\rm I.} \qquad   \epsilon > A  , \qquad z \in ( - \infty, +
\infty )  , \qquad R = \pm {1 \over \sqrt{A}} \,{\rm arcsinh}\,
\sqrt{{A \over \epsilon-A}}   \tanh  z   ; \nonumber
\\
{\rm II.} \qquad  \epsilon < A  , \qquad   \sinh^{2} z > {A -
\epsilon \over  \epsilon }   , \qquad
 R =
  \pm {1 \over \sqrt{A}}  \,{\rm arccosh}\, \sqrt{{A \over A - \epsilon}}   \tanh  z   .
%\label{6.12}
\end{gather*}

The left-hand side gives (two  dif\/ferent possibilities depending
on $(B^{2},A)$ relation)
\begin{gather*}
(a)\qquad  \left ( B^{2} > A , \, \mbox{f\/inite} \right ) \qquad
 L = {1 \over \sqrt{-a}} \arcsin { -2a  \cosh r  - b  \over
\sqrt{b^{2} - 4ac}}
 \; ,
\nonumber
\\
(b)\qquad  \left (  B^{2} \leq A ,  \, \mbox{inf\/inite} \right )
 \qquad L = {1 \over \sqrt{a}} \, \mbox{arccosh}\, { 2a   \cosh r  + b  \over
\sqrt{b^{2} - 4ac}}.
%\label{6.12'}
\end{gather*}
Therefore, the trajectories $F(r,z)$ (\ref{5.13d}) have the
form (four dif\/ferent cases)
\begin{gather*}
({\rm I}a) \qquad \epsilon > A  , \quad B^{2} >A   , \quad (I+B)^{2} >
B^{2}-A   , \quad
 z \in ( - \infty, + \infty )  , \quad  r \in ( r_{1} , r_{2} )   ,
\nonumber
\\
\phantom{({\rm I}a)} \qquad {1 \over \sqrt{-a}} \arcsin { -2a  \cosh  r - b
\over \sqrt{b^{2} - 4ac}}  =
  \pm {1 \over \sqrt{A}} \, \mbox{arcsinh}\left ( \sqrt{{A \over \epsilon-A}}   \tanh  z  \right)  ;
\nonumber
\\
({\rm II}a) \qquad \epsilon < A  ,\quad  B^{2} > A   ,
\quad(I+B)^{2} > B^{2}-A   , \quad
 \sinh^{2} z > {A \over  \epsilon }  -1   ,  \quad   r \in ( r_{1} , r_{2} )   ,
\nonumber
\\
\phantom{({\rm II}a)}{} \qquad {1 \over \sqrt{-a}} \, \mbox{arcsin}\, { -2a  \cosh  r - b
\over \sqrt{b^{2} - 4ac}}  =
  \pm {1 \over \sqrt{A}} \, \mbox{arccosh}\left(  \sqrt{{A \over A - \epsilon}}   \tanh  z \right)  ;
\nonumber
\\
({\rm I}b) \qquad \epsilon > A  , \quad B^{2} \leq A   , \quad z \in ( -
\infty, + \infty )  , \quad r \in ( r_{1} , \infty )   , \nonumber
\\
\phantom{({\rm I}b)}{}\qquad  {1 \over \sqrt{a}} \, \mbox{arccosh}\, { 2a   \cosh r  + b  \over
\sqrt{b^{2} - 4ac}} =
  \pm {1 \over \sqrt{A}} \, \mbox{arcsinh}\left( \sqrt{{A \over \epsilon-A}}   \tanh  z \right)  ;
\nonumber
\\
({\rm II}b) \qquad \epsilon < A  , \quad B^{2} \leq  A   , \quad
 \sinh^{2} z > {A  \over  \epsilon }  - 1 ,  \quad r \in ( r_{1} , \infty  )   ,
\nonumber
\\
\phantom{({\rm II}b)}{}\qquad  {1 \over \sqrt{a}} \, \mbox{arccosh}\, { 2a   \cosh r  + b  \over
\sqrt{b^{2} - 4ac}} =
  \pm {1 \over \sqrt{A}} \, \mbox{arccosh}\left( \sqrt{{A \over A - \epsilon}}   \tanh  z \right)  .
%\label{Classification}
\end{gather*}

\section[Trajectory equation  $F(r,\phi)=0$,
  the role of Lorentz $SO(3,1)$   transversal shifts in Lobachevsky space]{Trajectory equation  $\boldsymbol{F(r,\phi)=0}$,
  the role of Lorentz $\boldsymbol{SO(3,1)}$ \\  transversal shifts in Lobachevsky space}

Now, let us consider the trajectory equation  $F(r, \phi)$
(\ref{5.13e})
\begin{gather}
{ [   ( I +B)- B   \cosh  r    ]   dr \over \sinh
r  \sqrt{ A  \sinh^{2}r - [   (I +B) - B  \cosh  r
]^{2} }} = d \phi . \label{6.14a}
\end{gather}

With the help of  a new variable, the integral in the left-hand side reads
\begin{gather}
u = { (I + B ) \cosh  r -B  \over \sinh\  r }   ,
\qquad
L = - \int{  du  \over \sqrt{ [(A - B^{2})  + (I+B)^{2} ] - u^{2}
 }}   .
\label{6.14c}
\end{gather}
In connection to this  integral we must require
$
[(A - B^{2})  + (I+B)^{2} ] -  u^{2} > 0,$
which can be transformed to the form
\begin{gather*}
(A-B^{2})   \cosh^{2} r + 2B(I+B)   \cosh  r - A -
(I+B)^{2} > 0
\end{gather*}
or (see Section~\ref{section6})
$x = \cosh r$, $ax^{2} + b x +  c > 0$;
 therefore, all analysis given in Section~\ref{section6} is valid here as
well. In particular,  we should remember about necessary condition
\begin{gather*}
  b^{2} - 4ac  > 0  \qquad  \Longrightarrow
  \qquad
    (I+B)^{2} + (A -B^{2})  =  C^{2}> 0  .
\end{gather*}

Integral (\ref{6.14c}) equals to
\begin{gather*}
L =  \arccos  {u \over C }  =
  \arccos { (I + B )\cosh r -B  \over \sinh r
\sqrt{(I+B)^{2} + (A -B^{2})} }  ;
\end{gather*}
correspondingly, equation (\ref{6.14a}) gives
\begin{gather*}
 \arccos { (I + B )\cosh r -B  \over
\sqrt{(I+B)^{2} + (A -B^{2})}  \sinh r  } = \phi  ,
%\label{ad1}
\end{gather*}
or
\begin{gather}
   (I + B )\cosh r -
\sqrt{(I+B)^{2} + (A -B^{2})}   \sinh r  \cos \phi
= B  . \label{ad2}
\end{gather}

This is the most  general  form of the trajectory equation  $F(r, \phi) =0 $.
In particular, when $ (I+B)^{2} + (A -B^{2})   = 0 $,
 equation (\ref{ad2}) reads
as relationship def\/ining the motion of f\/ixed radius
$
   (I + B ) \cosh  r  = B$,  which coincides with (\ref{5.16}).

One may assume that equation (\ref{ad2}) when $ B^{2} >A$ will describe
a circle of f\/ixed radius  with shifted center in Lobachevsky
space. As we see below this assumption is true. Indeed, let us
introduce two coordinate systems in Lobachevsky space $H_{3}$
\begin{alignat*}{5}
& u_{1} = \cosh   z   \sinh  r \cos \phi   , \quad &&
u_{2} = \cosh   z   \sinh r \sin \phi   , \quad && u_{3}
= \sinh  z   , \quad &&   u_{0} = \cosh   z   \cosh
  r;&
\\
& u_{1} ' = \cosh   z'   \sinh  r' \cos \phi'  , \quad&&
u_{2}' = \cosh   z'   \sinh r'  \sin \phi '  , \quad&&
u_{3}' = \sinh  z '  , \quad  && u_{0}' = \cosh   z'
\cosh   r'   &
\end{alignat*}
 related by a  (Lorentz) shift in the plane (0--1)
\begin{gather}
\left | \begin{array}{c}
u'_{0} \\
u'_{1} \\
u'_{2} \\
u'_{3}
\end{array} \right | =
\left | \begin{array}{cccc}
\cosh  \beta  &  \sinh  \beta &  0  &  0 \\
\sinh  \beta &  \cosh  \beta & 0 & 0 \\
0  &  0  &  1  &  0 \\
0  &  0  &  0  &  1
\end{array} \right |
\left | \begin{array}{c}
u_{0} \\
u_{1} \\
u_{2} \\
u_{3}
\end{array} \right | .
\label{6.18a}
\end{gather}
Equations (\ref{6.18a})  result in
\begin{gather}
z' = z   ,  \qquad  \sinh  r'   \sin \phi ' =  \sinh
r   \sin \phi   , \nonumber
\\
\sinh  r'   \cos \phi' = \sinh \beta   \cosh  r +
\cosh  \beta   \sinh  r   \cos \phi   ,
\nonumber
\\
\cosh  r' = \cosh  \beta   \cosh  r + \sinh
\beta   \sinh  r  \cos \phi  . \label{6.18b}
\end{gather}
In particular,   $\beta$-shifted circle with f\/ixed value
$r' = r'_{0}$ will be described in coordinates $(r, \phi)$  by the
following equation
\begin{gather}
 \cosh \beta   \cosh  r + \sinh  \beta   \sinh  r  \cos \phi = \cosh  r'_{0}  .
\label{6.19a}
\end{gather}
This equation should be compared with the above equation
(\ref{ad2})
\begin{gather*}
   (I + B ) \cosh  r -
\sqrt{(I+B)^{2} - (A -B^{2})}   \sinh  r  \cos  \phi
= B  , %\label{6.19b}
\end{gather*}
 or (for simplicity, let $B$ and $I+B$ are  both positive)
\begin{gather}
   { (I + B ) \over \sqrt{ B^{2} -A} }  \cosh  r -
{ \sqrt{(I+B)^{2} - (A -B^{2})}  \over \sqrt{ B^{2} -A} }
\sinh  r  \cos  \phi    = {B   \over \sqrt{ B^{2} -A} }
  . \label{6.19c}
\end{gather}
Relations (\ref{6.19a})  and (\ref{6.19c}) coincide if
the parameter  $\beta$   and the radius of the shifted trajectory~$r_{0}'$
are  def\/ined according  to
\begin{gather*}
 \sinh  \beta = -
{ \sqrt{(I+B)^{2} - (A -B^{2})}  \over \sqrt{ B^{2} -A} }  ,
\qquad \cosh  r_{0}' = { B \over \sqrt{ B^{2} - A}}   .
%\label{6.22}
\end{gather*}

Let us turn back to the general equation for trajectories
$F(r,\phi)=0$ according to
 (\ref{ad2})
\begin{gather}
   (I + B ) \cosh  r -
\sqrt{(I+B)^{2} + (A -B^{2})}    \sinh  r   \cos \phi
= B   , \label{A1}
\end{gather}
 and describe its behavior under  (0--1)-shift
(\ref{6.18b})
\begin{gather*}
\sinh   r   \sin \phi  =  \sinh   r'   \sin \phi '  ,
\nonumber
\\
\sinh  r   \cos \phi = -\sinh \beta   \cosh  r ' +
\cosh   \beta   \sinh  r'   \cos \phi'   ,
\\
\cosh  r = \cosh \beta   \cosh  r' - \sinh\;
\beta  \sinh  r'  \cos \phi '  .
\end{gather*}

\noindent Trajectory equation $F(r, \phi)=0$ translated to
the coordinates $(r', \phi')$ looks
\begin{gather*}
   (I + B )  [ \cosh \beta  \cosh r' -
   \sinh \beta  \sinh r' \cos \phi '   ]
   \\
  \qquad{} -
\sqrt{(I+B)^{2} + (A -B^{2})}   [   -\sinh \beta
\cosh  r ' + \cosh   \beta   \sinh   r'   \cos
\phi' ]  = B   ,
\end{gather*}
or
\begin{gather}
\left [      \cosh  \beta    (I + B )
 + \sinh \beta   \sqrt{(I+B)^{2} + (A -B^{2})}    \right ]    \cosh  r '
 \nonumber
\\
  \qquad{} - \left [   \sinh  \beta   (I + B )
        +
 \cosh   \beta     \sqrt{(I+B)^{2} + (A -B^{2})}    \right ]   \sinh  r'   \cos \phi'    = B .
\label{A2}
\end{gather}

Comparing (\ref{A1}) and (\ref{A2}) we conclude that they are of
the same form if two simple combinations of parameters are transformed
by  means of a Lorentz shift\footnote{Below  in Section~\ref{section10} we will
show that a magnetic f\/ield turns to be invariant  under these
Lorentz shifts, so $B'=B$.}
\begin{gather}
I' + B =
 \cosh  \beta    (I + B )
 + \sinh  \beta   \sqrt{(I+B)^{2} + (A -B^{2})}   ,
 \nonumber
 \\
\sqrt{(I'+B)^{2} + (A' -B^{2})} = \sinh  \beta   (I + B )
+
 \cosh   \beta    \sqrt{(I+B)^{2} + (A -B^{2})}   .
 \label{A3}
 \end{gather}
These Lorentz shifts leave invariant the following
combination in the parametric space
\begin{gather*}
\mbox{inv} = (I+B)^{2} - \Big(\sqrt{(I+B)^{2} + (A -B^{2})}
\Big)^{2} = B^{2} - A   . %\label{A4}
\end{gather*}
This means that the Lorentz shifts vary in fact only
the parameter $I$, whereas $A'=A$.
It makes sense to introduce  new parameters $J$, $C$
\begin{gather*}
J = I + B   , \qquad C = + \sqrt{(I+B)^{2} + (A -B^{2})} =
\sqrt{I^{2} + 2IB + A} %\label{A5}
\end{gather*}
then equations (\ref{A3})  read
\begin{gather*}
J' =
J \cosh  \beta
 +C \sinh   \beta       , \qquad
C'  = J \sinh  \beta           +
 C \cosh  \beta
 %\label{A6}
 \end{gather*}
and invariant form of  the trajectory equation $F(r , \phi
)=0$ can be presented as
\begin{gather*}
   J \cosh r - C    \sinh r \cos \phi = B  ,
    %\label{A7}
\end{gather*}
in any other shifted reference frame it  must look as
(this is a direct result from invariance property $B'=B$ with
respect to transversal Lorentz shifts)
\begin{gather*}
   J ' \cosh r'  - C '    \sinh r'  \cos \phi ' = B  .
\end{gather*}
Correspondingly, the main invariant reads
\begin{gather*}
\mbox{inv} = J^{2} - C^{2} =   J^{\prime 2} - C^{\prime 2} = B^{2} - A   .
%\label{A8}
\end{gather*}

Depending on the sign of this invariant one can  reach the most
simple description by means of appropriate shifts

 1) $B^{2} - A  >0 $ ({\it finite motion})
\begin{gather*}
J_{0}^{2} = B^{2}- A   , \qquad C_{0} = 0   , \qquad
\mbox{trajectory}\quad    J _{0} \cosh  r  = B   ;
%\label{A9}
\end{gather*}

2) $B^{2} - A < 0 $ ({\it infinite motion})
\begin{gather*}
J_{0} =  0   , \quad C^{2}_{0} = A - B^{2}     \qquad
 \mbox{trajectory}\quad
   - C_{0}     \sinh  r \cos \phi  = B  .
%\label{A9}
\end{gather*}

One special case exists, see (\ref{addition-1}), (\ref{addition-2}).

 3) $B^{2} = A$ ({\it infinite motion})
\begin{gather*}
J = I + B , \qquad C = I + B   , \qquad \mbox{trajectory }\quad
  \cosh  r -  \sinh  r \cos \phi = {B \over  I + B}  .
 %\label{A10}
\end{gather*}

By symmetry reasons, Lorentzian shifts of the type (0--2) will
manifest themselves in the same manner.

\section[Lorentzian   shifts   and  symmetry  of a  magnetic field  in $H_{3}$]{Lorentzian   shifts   and  symmetry  of a  magnetic f\/ield  in $\boldsymbol{H_{3}}$}\label{section9}

Let us turn again to a pair of coordinate systems in space
$H_{3}$
\begin{gather*}
u_{1} = \cosh   z   \sinh  r \cos \phi , \qquad
u_{2} = \cosh   z   \sinh r \sin \phi   , \nonumber
\\
u_{3} = \sinh  z   , \qquad  u_{0} = \cosh   z
\cosh   r  ; \nonumber
\\
u_{1} ' = \cosh   z'   \sinh  r' \cos \phi'   , \qquad
u_{2}' = \cosh   z'   \sinh r'  \sin \phi '   ,
\nonumber
\\
u_{3}' = \sinh  z '  , \qquad  u_{0}' = \cosh   z'
\cosh   r'  , %\label{12.1b}
\end{gather*}
related by the shift  (0--1)
\begin{gather*}
\left | \begin{array}{c}
u'_{0} \\
u'_{1} \\
u'_{2} \\
u'_{3}
\end{array} \right | =
\left | \begin{array}{cccc}
\mbox{ cosh}\; \beta  &  \sinh\; \beta &  0  &  0 \\
\sinh\; \beta &  \cosh\; \beta & 0 & 0 \\
0  &  0  &  1  &  0 \\
0  &  0  &  0  &  1
\end{array} \right |
\left | \begin{array}{c}
u_{0} \\
u_{1} \\
u_{2} \\
u_{3}
\end{array} \right |  ,
\end{gather*}
or in cylindric coordinates (direct and inverse
formulas)
\begin{gather*}
z' = z   , \qquad \sinh   r'   \sin \phi ' =  \sinh
r   \sin \phi   , \nonumber
\\
\sinh  r'  \cos \phi' = \sinh \beta   \cosh r +
\cosh   \beta   \sinh  r   \cos \phi  ,
\nonumber
\\
\cosh  r' = \cosh  \beta   \cosh  r + \sinh
\beta  \sinh  r \cos \phi  ; \nonumber
\\
z = z '  , \qquad \sinh  r   \sin \phi  =  \sinh   r'
 \sin \phi'   , \nonumber
\\
\sinh  r   \cos \phi =  - \sinh \beta   \cosh  r'
+ \cosh  \beta   \sinh  r '   \cos \phi '  ,
\nonumber
\\
\cosh  r = \cosh  \beta   \cosh  r'  - \sinh
\beta   \sinh  r'   \cos \phi'   . %\label{12.2c}
\end{gather*}

With respect to that  coordinate change  $(r, \phi)
\Longrightarrow (r', \phi')$, the uniform magnetic f\/ield transforms
according to
\begin{gather*}
F_{\phi' r'} = {\partial x^{\alpha} \over \partial \phi '}
{\partial x^{\beta} \over \partial r'} F_{\alpha \beta}= \left(
{\partial \phi \over \partial \phi '}  {\partial r  \over \partial
r'} -{\partial r \over \partial \phi ' }  {\partial \phi \over
\partial r'}  \right) F_{\phi r}     , \qquad F_{\phi r} = B
\sinh  r  , %\nonumber \label{12.3a}
\end{gather*}
so that the magnetic f\/ield transforms by means of  the
Jacobian
\begin{gather*}
F_{\phi' r'} = J  F_{\phi r}   , \qquad J = \left |
\begin{array}{cc}
{\partial r \over \partial r' }   &  {\partial r \over \partial \phi ' }  \vspace{2mm}\\
{\partial \phi  \over \partial r' }  & {\partial \phi  \over
\partial \phi' }
\end{array} \right | , \qquad   F_{\phi r}  = B    \sinh  r   .
%\label{12.3b}
\end{gather*}

It is convenient to represent the coordinate transformation in the
form
\begin{gather*}
\phi =  \arctan   \left (    { \sinh  r'    \sin \phi' \over - \sinh  \beta   \cosh  r' + \cosh  \beta
  \sinh r' \cos \phi' } \right )
 = \arctan   A   ,
\nonumber
\\
r = \mbox{arccosh}  \left (    \cosh  \beta   \cosh  r'
- \sinh  \beta   \sinh  r'   \cos \phi'   \right  )=
\mbox{arccosh}\, B   ; %\label{12.4a}
\end{gather*}
correspondingly  the Jacobian  looks
\begin{gather*}
J = {1 \over \sqrt{B^{2}-1}}  {1 \over 1 +A^{2}}   \left (
{\partial B \over \partial r' }  {\partial A \over \partial \phi
'} - {\partial B \over \partial \phi' }   {\partial A \over
\partial r '}   \right )  . %\label{12.4b}
\end{gather*}
Taking into account the  identity
\begin{gather*}
{1 \over \sqrt{B^{2}-1}}  {1 \over 1 +A^{2}}= {1 \over
\sqrt{\cosh^{2} r -1}} {1 \over 1 + \tan^{2}  \phi}=
{\cos^{2} \phi \over  \sinh   r}   ,
\end{gather*}
and the  formulas
\begin{gather*}
{\partial B \over \partial r' } = {\partial  \over \partial r' }
  (  \cosh  \beta   \cosh  r'  - \sinh  \beta   \sinh  r'   \cos \phi' )=
\cosh  \beta  \sinh  r'  - \sinh  \beta
\cosh  r'   \cos \phi'   , \\
{\partial B \over \partial \phi' } = {\partial  \over \partial
\phi ' }   (  \cosh  \beta   \cosh  r'  - \sinh
\beta   \sinh  r'   \cos \phi' )=
 \sinh  \beta   \sinh  r'   \sin \phi'   ,
\\
{\partial A \over \partial \phi' }= {\partial  \over \partial
\phi' }   \left (    { \sinh  r'    \sin \phi ' \over -
\sinh  \beta   \cosh  r' + \cosh  \beta
\sinh r' \cos \phi' } \right ) \\
\phantom{{\partial A \over \partial \phi' }}{}
=
 { \sinh  r'    ( - \sinh  \beta   \cosh  r' \cos \phi' +
 \cosh  \beta   \sinh  r' )  \over
( - \sinh  \beta   \cosh  r' + \cosh  \beta
\sinh r' \cos \phi' )^{2} }
\\
\phantom{{\partial A \over \partial \phi' }}{}
= { \sinh  r'    ( -
\sinh  \beta   \cosh  r' \cos \phi' +
 \cosh  \beta   \sinh  r' )  \over  \sinh^{2} r   \cos^{2} \phi }   ,
\\
{\partial A \over \partial r' }= {\partial  \over \partial r' }
\left (    { \sinh  r'    \sin \phi ' \over - \sinh
\beta   \cosh  r' + \cosh  \beta   \sinh r' \cos
\phi' } \right ) \\
\phantom{{\partial A \over \partial r' }}{}
= { - \sinh  \beta   \sin \phi '
 \over
( - \sinh  \beta   \cosh  r' + \cosh  \beta
\sinh r' \cos \phi' )^{2} }= { - \sinh  \beta   \sin
\phi '
 \over   \sinh^{2} r   \cos^{2} \phi }   ,
\end{gather*}
 for the Jacobian we get
\begin{gather*}
J= { \sinh  r'   \over  \sinh   r}
{1 \over \sinh^{2} r }
 \big[   ( \cosh
\beta   \sinh  r'  - \sinh  \beta   \cosh  r'
\cos \phi' ) \\
 \phantom{J=}{}\times  ( - \sinh  \beta   \cosh  r' \cos \phi'
   +
 \cosh  \beta   \sinh  r' )  +
 \sinh^{2} \beta    \sin ^{2}\phi' \big]  .
% \label{12.5c}
 \end{gather*}
It is matter of simple calculation to verify that
the denominator  and numerator
\begin{gather*}
\sinh^{2}r =   \cosh^{2} r -1 =
 (\cosh  \beta   \cosh  r'  - \sinh  \beta   \sinh  r'   \cos \phi')^{2}  -1
\\
\phantom{\sinh^{2}r}{}=\cosh^{2} \beta   \cosh^{2} r'   - 2 \cosh  \beta
\sinh  \beta
  \cosh  r'  \sinh  r'   \cos \phi'  +
 \sinh^{2}\beta   \sinh^{2}r'  \cos ^{2}\phi' - 1     ,
\\
 ( \cosh  \beta   \sinh  r'  - \sinh  \beta   \cosh  r'   \cos \phi' )
  ( - \sinh  \beta   \cosh  r' \cos \phi' +
 \cosh  \beta   \sinh  r' )
 \\
\qquad{} +
 \sinh^{2} \beta   \sin ^{2}\phi'
   =
 -2 \cosh \beta   \sinh \beta  \cosh r'   \sinh r'  \cos \phi'
\\
\qquad{}+
 \cosh^{2}\beta  \sinh^{2} r' +  \sinh^{2}\beta  \cosh^{2} r'  \cos^{2}\phi' +
 \sinh^{2} \beta \sin^{2} \phi '
 \end{gather*}
are equal to each other. Thus, the  Jacobian of the
shift  (0--1)  in hyperbolic space is
\begin{gather*}
J =  { \sinh  r'   \over  \sinh   r}
 %\label{12.6a}
 \end{gather*}
and therefore this shift  leaves invariant the uniform
magnetic f\/ield under consideration
\begin{gather*}
F_{\phi' r'} = J  F_{\phi r}   , \qquad
  F_{\phi r}  = B    \sinh  r   , \qquad F_{\phi' r'} =  B    \sinh  r'  .
%\label{12.6b}
\end{gather*}

By symmetry reason, we can conclude the same result for the shifts of
the type
(0--2).
However, in that sense shifts of the type  (0--3) behave
dif\/ferently. Indeed,
\begin{gather}
\left | \begin{array}{c}
u'_{0} \\
u'_{1} \\
u'_{2} \\
u'_{3}
\end{array} \right | =
\left | \begin{array}{cccc}
\mbox{ cosh}\; \beta  &  0 &  0  &  \sinh\; \beta  \\
0  &  1 & 0 & 0 \\
0  &  0  &  1  &  0 \\
\sinh\; \beta   &  0  &  0  &  \cosh\; \beta
\end{array} \right |
\left | \begin{array}{c}
u_{0} \\
u_{1} \\
u_{2} \\
u_{3}
\end{array} \right | .
\label{12.7a}
\end{gather}
Equation (\ref{12.7a}) leads to relation  between $(r,z)$
and $(r', z')$
\begin{gather*}
\cosh  z'   \cosh  r' = \cosh \beta   \cosh
z   \cosh  r +
 \sinh \beta   \sinh  z   ,
\nonumber
\\
 \sinh  z' = \sinh \beta    \cosh  z   \cosh  r +
 \cosh  \beta   \sinh  z   ,
\nonumber\\
\cosh  z'  \sinh  r' = \cosh z   \sinh r , \qquad \phi ' = \phi   . %\label{12.8a}
\end{gather*}
Electromagnetic f\/ield is transformed according to
\begin{gather*}
F_{\alpha ' \beta'} = \left( {\partial \phi \over \partial x^{\prime \alpha}}
{\partial r \over \partial x^{\prime \beta}} - {\partial r \over
\partial x^{\prime \alpha}}  {\partial \phi  \over \partial x^{\prime \beta}}
\right) F_{ \phi r} \quad \Longrightarrow \quad F_{\phi' r'} = {\partial r
\over \partial r'}   F_{\phi r}  , \qquad F_{\phi' z'} = {\partial
r \over \partial z'}   F_{\phi r}   ,%\label{12.8b}
\end{gather*}
so  the uniform magnetic f\/ield in the
space  $H_{3}$ is not invariant with respect to
 the shifts (0--3).

One may describe electromagnetic f\/ield in terms of 4-potential,
and  the rule to transform the f\/ield  with respect to the shift
(0--1) looks
\begin{gather*}
 A_{\phi} =  - B (\cosh  r -1 )\quad   \Longrightarrow \quad
 A'_{\phi'} = {\partial \phi   \over \partial \phi' }    A_{\phi}   , \qquad
A'_{r'} = {\partial \phi   \over  \partial r '}    A_{\phi} .
% \label{12.9}
 \end{gather*}
In f\/lat space,  the  shift (0--1) generates a  def\/inite
gauge transformation
\begin{gather*}
{\bf A}({\bf r})  = {1 \over 2}    {\bf B} \times {\bf r}, \qquad
{\bf r} ' = {\bf r} + {\bf b}   , \nonumber
\\
{\bf A}'({\bf r}') = {1 \over 2} {\bf B} \times {\bf r}' - {1
\over 2} {\bf B} \times {\bf b} = {1 \over 2} {\bf B} \times {\bf
r}' + \nabla _{{\bf r}'}    \Lambda (x', y', z')   , \nonumber
\\
{\bf B} = (0,0,B) , \qquad {\bf b } = (b, 0,0)   , \qquad
\Lambda  (x', y', z')  = -  {bB \over 2}   y '    .
%\label{12.10}
\end{gather*}
By analogy reason, one could expect something similar in
the case of Lobachevsky space as well
\begin{gather}
 A'_{\phi'} = {\partial \phi   \over \partial \phi' }    A_{\phi}  =
 - B  (\cosh  r' -1 ) +  {\partial \over \partial \phi'}   \Lambda  , \qquad
A'_{r'} = {\partial \phi   \over  \partial r '}   A_{\phi}  =
{\partial \over \partial r '}  \Lambda  . \label{12.11}
\end{gather}
It is indeed so. Let us demonstrate this. Accounting for
two formulas
\begin{gather*}
{\partial \phi \over \partial \phi '} = {1 \over 1 + A^{2} }
{\partial A \over \partial \phi '} = { \sinh  r'    ( -
\sinh  \beta   \cosh  r' \cos \phi' +
 \cosh  \beta   \sinh  r' )  \over  \sinh^{2} r  }  ,
\\
{\partial \phi \over \partial r '} = {1 \over 1 + A^{2} }
{\partial A \over \partial r '} = { - \sinh  \beta  \sin
\phi '
 \over   \sinh^{2} r   }   ,
\end{gather*}
we conclude that the gauge function  $\Lambda$ in
(\ref{12.11})  is def\/ined by its partial derivatives in accordance
with
\begin{gather}
{\partial \Lambda \over \partial r'} = B   {  \sinh  \beta
  \sin \phi '
 \over    1 + \cosh  \beta   \cosh  r'  - \sinh  \beta   \sinh  r'   \cos \phi'      }   ,
\nonumber
\\
{\partial \Lambda \over \partial \phi'} = B (\cosh  r' -1) -
B  { \sinh  r'    (  \cosh  \beta   \sinh  r' -
\sinh  \beta   \cosh  r' \cos \phi'
  )  \over
 1 + \cosh  \beta   \cosh  r'  - \sinh  \beta  \sinh r'   \cos \phi'     }
\nonumber
\\
\phantom{{\partial \Lambda \over \partial \phi'}}{}
= B  { (\cosh  r' -1) (1 - \cosh  \beta) + \sinh
\beta   \sinh r' \cos \phi '\over
 1 + \cosh  \beta   \cosh  r'  - \sinh  \beta   \sinh  r'   \cos \phi'     }   .
\label{12.14}
\end{gather}
The integrability  condition
$\partial ^{2}  \Lambda  / \partial \phi' \partial  r '=
\partial ^{2} \Lambda /  \partial  r' \partial  \phi '$ in an explicit form looks
\begin{gather*}
{\partial \over \partial \phi '}   {  \sinh  \beta   \sin
\phi '
 \over    1 + \cosh  \beta   \cosh  r'  - \sinh  \beta   \sinh  r'   \cos \phi'      }
 \nonumber
 \\
 \qquad {} =
 {\partial \over  \partial r'}
  { (\cosh  r' -1) (1 - \cosh  \beta) + \sinh  \beta  \sinh r' \cos \phi '\over
 1 + \cosh  \beta   \cosh  r'  - \sinh  \beta   \sinh  r'   \cos \phi'     }   .
%\label{12.15}
\end{gather*}

It is   the matter of simple direct calculation to verify it. Now
we are to f\/ind an explicit form of the gauge function $\Lambda$. For
better understanding, it is  helpful  f\/irst to consider a similar
problem in f\/lat space
\begin{gather*}
\mbox{space}\   E_{3}, \qquad  {\partial \Lambda \over \partial r'} = - B  {   \alpha  \sin
\phi '
 \over    2     }  , \qquad
{\partial \Lambda \over \partial \phi'} =  -B{   \alpha r' \cos
\phi '\over  2    } . %\label{12.16a}
\end{gather*}
Integrating the f\/irst equation we obtain
\begin{gather*}
\Lambda = -{B \alpha \over 2 } \sin \phi '  r' + \lambda (\phi')
\end{gather*}
and further from the second equation we  derive
\begin{gather*}
 -{B \alpha \over 2 } \cos \phi '   r' + {d  \over d \phi } \lambda (\phi') =
 -B  {   \alpha r' \cos \phi '\over  2    } \qquad   \Longrightarrow   \qquad  \lambda (\phi ') = \lambda   .
\end{gather*}
Therefore, the  gauge function is
\begin{gather*}
\Lambda ( r' , \phi ' ) = -{B \alpha \over 2 } \sin \phi '   r' +
\lambda = - {B \alpha \over 2 }  y ' + \lambda   .
%\label{12.16b}
\end{gather*}

The similar problem in space $H_{3}$ should be considered by the
same scheme. Let us integrate  the f\/irst equation in
(\ref{12.14})
\begin{gather*}
 \Lambda  =  B   \sinh  \beta   \sin \phi '   \int
   {  dr'
 \over    1 + \cosh  \beta   \cosh  r'  - \sinh  \beta   \sinh  r'   \cos \phi'      }
 + \lambda ( \phi ')    .
%\label{12.17a}
\end{gather*}
Introducing the notation
$
\cosh  \beta = c$, $\sinh  \beta = s$, $c^{2} - s^{2} = 1 $,  and new variable
\begin{gather*}
\tanh   {r' \over 2} = y   , \qquad dr' = 2   \cosh^{2}
r'   dy   , \nonumber\\
 {  dr'
 \over    1 + \cosh  \beta   \cosh r'  - \sinh  \beta   \sinh  r'   \cos \phi'      }
%\label{12.17b}
\\
\quad =  {  2   \cosh^{2} r'  dy
 \over    1 + c  ( \cosh^{2}{ r'\over 2} + \sinh^{2}{ r'\over 2} )
  - s   2 \sinh { r'\over 2}   \cosh { r'\over 2}    \cos \phi'      }=
  {  2   dy
 \over    y^{2}(c-1)
  - 2 y   s   \cos \phi'    + c + 1     }   ,\nonumber
\end{gather*}
for $\Lambda$  we get
\begin{gather*}
\Lambda = \lambda (\phi' ) + 2B \arctan  { (c-1) y - s
\cos \phi ' \over s \sin \phi ' }   ,
\end{gather*}
from whence it follows
\begin{gather*}
\Lambda = \lambda (\phi') + 2B \arctan { (c-1)
(\cosh  r' -1)  - s   \sinh r'  \cos \phi ' \over
s  \sinh r'    \sin \phi '}  .
% \label{12.17c}
 \end{gather*}
Now we are to calculate
\begin{gather*}
{\partial \Lambda \over \partial \phi '} = {d \lambda \over d \phi
'} +2B {s^{2}  \sinh^{2}r'    \sin^{2} \phi ' - [(c-1) (\cosh  r' -1)  - s   \sinh r'  \cos \phi
' ] s  \sinh r' \cos \phi'
 \over s^{2}  \sinh^{2}r'    \sin^{2} \phi ' +
 [  (c-1) (\cosh  r' -1)  - s \;  \sinh\; r' \; \cos \phi '\; ]^{2} }
\\
\phantom{\partial \Lambda \over \partial \phi '}{}{} = {d \lambda \over d \phi '} +2B {(c+1)  (\cosh r'+1) -
   s \sinh r' \cos \phi '
 \over (c+1)  (\cosh r' +1)   +
  (c-1) (\cosh  r' -1)  - 2 s   \sinh  r'   \cos \phi ' }
\end{gather*}
and f\/inally
\begin{gather*}
{\partial \Lambda \over \partial \phi '} = {d \lambda \over d \phi
'} +B  { (\cosh  r'+1) (c+1)   -
   s   \sinh  r' \cos \phi '
 \over
c   \cosh   r'  + 1 -  s   \sinh  r'   \cos \phi
' }  . %\label{12.18a}
\end{gather*}
After substituting into
the second equation in
 (\ref{12.14}) the expression for $\Lambda$   gives
\begin{gather*}
{d \lambda \over d \phi '} =
  B { (\cosh  r' -1) (-c+1) + s   \sinh r' \cos \phi '\over
 1 + c   \cosh  r'  - s   \sinh  r'  \cos \phi'     }
 \nonumber
 \\
\phantom{{d \lambda \over d \phi '} =}{}  +
B { (\cosh  r'+1) (-c-1)    +
   s   \sinh  r' \cos \phi
 \over
 1 +  c   \cosh   r'  -  s    \sinh  r'  \cos \phi ' } = -2B  .
%\label{12.18b}
\end{gather*}

Therefore,
\begin{gather*}
\lambda (\phi)  = - 2B \phi + \lambda _{0}
 %\label{12.19}
\end{gather*}
and the gauge function $\Lambda (r', \phi')$  is found as
\begin{gather*}
\Lambda (r', \phi')=  + 2B \arctan { (c-1) (\cosh
r' -1)  - s   \sinh r'  \cos \phi ' \over s
\sinh r'    \sin \phi '}
 - 2B \phi' + \lambda _{0}
%\label{12.20}
\end{gather*}
remembering that $ c =\cosh  \beta$, $s =
\sinh  \beta .$

%\begin{center}
%{\bf PART II. PARTICLE IN THE MODEL $S_{3}$ }
%\end{center}

\section[Particle in a magnetic field, spherical Riemann model  $S_{3}$]{Particle in a magnetic f\/ield, spherical Riemann model  $\boldsymbol{S_{3}}$}\label{section10}

In \cite{Olevsky}  under  number  XI  we see the following
system of coordinates in spherical space $S_{3}$
\begin{gather*}
dS^{2} = c^{2} dt^{2} - \rho^{2}  [   \cos^{2} z ( d r^{2} +
\sin^{2} r   d \phi^{2} ) + dz^{2}  ]  , \nonumber
\\
z \in [-\pi /2  , + \pi /2 ]  , \qquad r \in [0, + \pi ]   ,
\qquad \phi \in [0, 2 \pi ]   , \nonumber
\\
u^{1} = \cos z   \sin  r \cos \phi   , \qquad  u^{2} = \cos z   \sin
r \sin \phi   ,\qquad u^{3} = \sin  z   ,   \nonumber
\\
  u^{0} = \cos   z   \cos r  , \qquad  \big(u^{0}\big)^{2} + \big(u^{1}\big)^{2} + \big(u^{2}\big)^{2} + \big(u^{3}\big)^{2} = 1   .
%\label{7.2}
\end{gather*}
In these coordinates, let us introduce a magnetic f\/ield
\begin{gather*}
 A_{\phi} = -2B \sin^{2} {r \over 2}  = B  ( \cos  r -1 )  , \qquad
 F_{\phi r} =
\partial_{\phi}A_{r} -  \partial_{r} A_{\phi} =  B    \sin  r   ,
%\label{7.3b}
\end{gather*}
which satisf\/ies the Maxwell equations in $S_{3}$
\begin{gather*}
 {1 \over
\cos^{2} z   \sin r } {\partial \over
\partial r }  \cos^{2} z   \sin  r   \left( {1 \over
\cos^{4}z    \sin ^{2}{r} } \right) B \sin  r  \equiv 0   .
\end{gather*}
The Christof\/fel symbols read
\begin{gather*}
\Gamma^{r}_{\;\;jk } = \left | \begin{array}{ccc}
0 & 0 & -\tan z \\
0 & - \sin r \cos  r & 0 \\
- \tan z  & 0 & 0
\end{array} \right |   , \qquad
\Gamma^{\phi}_{\;\;jk } = \left | \begin{array}{ccc}
0 &  \cot  r  & 0\\
\cot  r  & 0 &- \tan z \\
0 & -\tan  z  & 0
\end{array} \right |  ,
\\
\Gamma^{z}_{\;\;jk } = \left | \begin{array}{ccc}
\sin z \cos  z  &  0  & 0\\
0  &  \sin  z   \cos z \sin^{2} r & 0 \\
0 & 0 & 0
\end{array} \right |   .
\end{gather*}
The non-relativistic equations of motion (\ref{1.7b}) in
coordinates  $( r , \phi , z) $
 in a magnetic f\/ield  look
\begin{gather}
  {d V^{r}  \over d t   } -
2 \tan  z   V^{r}  V^{z} - \sin r \cos r   V^{\phi }
V^{\phi} =
  B   { \sin  r    \over \cos^{2} z }  V^{\phi}       ,
\nonumber
\\
{d  V^{\phi}  \over d t  } + 2 \cot r  V^{\phi }
V^{r}  -   2 \tan z   V^{\phi } V^{z} =
 - B   {1   \over \cos^{2} z \sin  r}    V^{r}    ,
\nonumber
\\
{d  V^{z}  \over d t  }  + \sin z \cos z    V^{r}  V^{r} + \sin
z \cos z  \sin^{2} r  V^{\phi}  V^{\phi}=  0    . \label{7.9}
\end{gather}
The last equation in  (\ref{7.9})  points that along the
axis $z$ the ef\/fective attractive force acts to the center $z=0$.

\section{Simplest solutions in spherical space}\label{section11}

Let us search for solutions with a f\/ixed radius $ r = r_{0}   $; equations
(\ref{7.9}) give
\begin{gather*}
    V^{\phi }  = -   { B  \over \cos  r_{0}}   { 1  \over \cos ^{2} z }        ,
\qquad
{d    \over d t  }  \left( -   { B  \over  \cos r_{0}}   { 1  \over
\cos ^{2} z } \right) -
    2 \tan  z     \left(  -   { B  \over \cos r_{0}}   { 1  \over \cos^{2} z }\right)  V^{z} = 0    ,
\nonumber
\\
{d  V^{z}  \over d t  }  =  -\big(   \tan^{2} r_{0}   B^{2}
\big)   {  \sin z  \over \cos^{3} z },
    %\label{8.1}
    \end{gather*}
 where the second equation is an identity $0 \equiv 0$.
With notation
$
\alpha = -    B  /  \cos  r_{0}$, $A = B^{2}  \tan^{2} r_{0} $,
 the problem reduces to
\begin{gather}
    {d \phi \over dt}   =  { \alpha  \over \cos^{2} z }        , \qquad
{d  V^{z}  \over d t  }  =  - A   {  \sin  z  \over \cos^{3} z }
 . \label{8.2}
\end{gather}
The second equation gives
\begin{gather}
d  ( V^{z})^{2}  =  A     d \left(  -{1 \over \cos^{2} z}  \right) \qquad
\Longrightarrow \qquad
 \left({dz \over dt}\right)^{2} = -{A \over \cos^{2} z } +
\epsilon, \label{8.3}
\end{gather}
the constant  $\epsilon$   will be related  to the
squared velocity. First, let  $A \neq  \epsilon$, then equation~(\ref{8.3}) gives (in contrast to Lobachevsky model, now only one
possibility is realized: $\epsilon > A$)
\begin{gather*}
 {  d   \sin  z  \over
\sqrt{ \epsilon  (1 - \sin^{2} z )   - A }  } = dt  .
%\label{8.4}
\end{gather*}
Two dif\/ferent signs $(\pm)$ correspond to the motion of
the particle in opposite directions along the axis
 $z$.  The calculations below are evident  (let   $t_{0}=0$)
\begin{gather*}
A \neq  \epsilon   , \qquad \sin  z  (t) =     \pm   {
\sqrt{\epsilon -A} \over \sqrt{\epsilon}}   \sin \sqrt{\epsilon}
   t   . %\label{8.5}
\end{gather*}

Let us examine a special case when   $\epsilon = A$,  equation
(\ref{8.3})  becomes
\begin{gather}
\left({dz \over dt}\right)^{2} = - \epsilon \tan ^{2} z.
\label{8.6}
\end{gather}
Equation (\ref{8.6}) has only a   trivial solution
\begin{gather*}
z (t) = 0   , \qquad \mbox{so that} \qquad  \phi (t) = \phi
_{0} + \alpha   t   , \qquad \alpha = -   { B  \over \cosh
r_{0}}, %\label{8.7}
\end{gather*}
it  corresponds to the rotation with constant angular
velocity around  the circle  $r=r_{0}$ in the absence of any  motion
along the axis~$z$.

Now we are to turn  to the f\/irst equation in   (\ref{8.2}) and
f\/ind   $\phi (t)$
\begin{gather*}
A \neq \epsilon  , \qquad \phi     =  \alpha    \int  {   dt
\over \cos^{2} z }  = \alpha  \int {  d t    \over
\cos^{2} \sqrt{\epsilon}  t +    {A  \over \epsilon }  \sin^{2}
\sqrt{\epsilon}   t  }    ,
\end{gather*}
so that
\begin{gather}
A \neq \epsilon  , \qquad \phi    =   {  \alpha  \over \sqrt{ A}
} \arctan  \left( \sqrt{{A \over \epsilon}} \tan
\sqrt{\epsilon}  t \right)   . \label{8.8a}
\end{gather}

Thus, we construct the following solution
\begin{gather*}
r= r_{0} = \mbox{const}   , \qquad A \neq  \epsilon   ,
\nonumber
\\
\sin  z  (t) =     \pm   { \sqrt{\epsilon -A} \over
\sqrt{\epsilon}}   \sin \sqrt{\epsilon}    t    , \qquad \phi
=   {  \alpha  \over \sqrt{ A} } \arctan  \left( \sqrt{{A
\over \epsilon}} \tan  \sqrt{\epsilon}  t \right) .
%\label{8.9}
\end{gather*}
Distinctive feature of the motion is  its periodicity
and  closeness of corresponding trajectories. The period
$T$ is determined by the energy $\epsilon$
\begin{gather*}
T = { \pi \over \sqrt{\epsilon} } , \qquad  \mbox{or in usual
units} \qquad  T =
  { \pi   \rho  \over V }   .
%\label{8.10}
\end{gather*}

\section[Particle in a magnetic field and Lagrange formalism in $S_{3}$]{Particle in a magnetic f\/ield and Lagrange formalism in $\boldsymbol{S_{3}}$}\label{section12}

Let us  consider the problem of a particle in a magnetic f\/ield in
$S_{3}$ on the base of the Lagrange function
\begin{gather*}
L = {1 \over 2}   \big(   \cos^{2}z   V^{r}  V^{r}  + \cos^{2}z
\sin^{2}r  V^{\phi}  V^{\phi}
 +   V^{z} V^{z}   \big)  -   B   ( \cos r -1 ) V^{\phi}     .
%\label{10.1}
\end{gather*}
 Euler--Lagrange equations
 in the explicit  form are
\begin{gather*}
{d \over dt}   \cos^{2} z V^{r}  =   \cos^{2} z \sin  r \cos  r
V^{\phi} V^{\phi} +  B   \; \sin  r  V^{\phi}   , \nonumber
\\
{d \over dt}  \big[   \cos^{2}z \sin^{2}r  V^{\phi}  -   B   (
\cos r -1 )  \big]  = 0  , \nonumber
\\
{d \over dt}   V^{z} = - \cos  z  \sin  z    (  V^{r}  V^{r}  +
\sin^{2}r  V^{\phi}  V^{\phi} )  , %\label{10.2}
\end{gather*}
or  dif\/ferently
\begin{gather*}
  {d V^{r}  \over d t   } -
2 \tan  z  V^{r}  V^{z} - \sin  r \cos  r  V^{\phi }
V^{\phi} =
 B   { \sin r    \over \cos^{2} z }  V^{\phi}     ,
\nonumber
\\
{d \over dt}  \big[   \cos^{2}z\; \sin^{2}r\; V^{\phi}  -   B  (
\cos r -1 )  \big]  = 0   , \nonumber
\\
{d \over dt}   V^{z} =- \cos z \sin  z    (  V^{r}  V^{r}  +
\sin^{2}r  V^{\phi}  V^{\phi} )   , %\label{10.3}
\end{gather*}
which coincide with equations (\ref{7.9}). Two integrals of
motion in $S_{3}$ are known
\begin{gather*}
I =   \cos^{2}z  \sin^{2}r  V^{\phi}  -
 B   ( \cos  r -1 )     , \qquad
  \epsilon = \cos^{2}z   ( V^{r}  V^{r}  +  \sin^{2}r  V^{\phi}  V^{\phi}  )  +   V^{z} V^{z},
%\label{10.5}
\end{gather*}
the third one can be constructed as  follows
 \begin{gather*}
 A = \cos^{2} z   \left[    \epsilon - \left({d z \over dt}\right)^{2}  \right]
=  \cos^{4} z   (       V^{r}  V^{r}  +   \sin^{2}r  V^{\phi}
V^{\phi}   )   . %\label{10.6}
\end{gather*}

With the help of tree integrals of motion one can readily
transform the  problem under consideration to calculating several
integrals
\begin{gather}
 {d \phi \over d t} =  {1 \over   \cos ^{2} z  }  { I + B   ( \cos r -1 ) \over  \sin^{2} r }  ,
\label{10.13a}
\\
{d r \over dt} = \pm   {1 \over \cos^{2} z  }  \sqrt { A - {
[   I + B   ( \cos r -1)   ] ^{2} \over
  \sin^{2} r }  } ,
\label{10.13b}
\\
{dz \over dt} =  \pm   {1 \over \cos z}    \sqrt{ \epsilon
\cos ^{2} z - A}   ,
\label{10.13c}
\\
{  \sin  r  d r \over  \sqrt   { A   \sin ^{2} r  -
  (I + B   \cos  r - B)^{2}  } } =  \pm  {1 \over \cos   z   }
  { dz  \over \sqrt{ \epsilon  \cos^{2} z - A}}     ,
\label{10.13d}
\\
{ [  I + B( \cos r -1)   ]  dr \over \sin  r  \sqrt{ A
\sin^{2}r - [   I + B ( \cos  r -1)  ]^{2} }} = d \phi   .
\label{10.13e}
\end{gather}

For the most simple case when  $r = r_{0} = \mbox{const}$,  these
relations  give
\begin{gather}
 {d \phi \over d t} = {\alpha  \over  \cosh^{2} z  }
  , \qquad
 {dz \over dt} =  \pm   {1 \over \cos z}    \sqrt{
\epsilon\; \cos ^{2} z - A}    , \nonumber
\\
 \alpha =   { I + B    ( \cos r_{0} -1 ) \over   \sin^{2} r_{0} }    , \qquad
 A =     { [  I + B   ( \cos r_{0} -1)  ]^{2} \over
  \sin ^{2} r_{0} }     .
\label{10.13}
\end{gather}
In Section~\ref{section11} for $\alpha$ and  $A$   we had
other expressions
\begin{gather}
\alpha = -   { B  \over \cos  r_{0}}   , \qquad A = (
\tan^{2} r_{0}    B^{2}  ) , \label{10.14}
\end{gather}
which were equivalent to the present ones. Indeed, from
the identity  $\alpha = \alpha$ we get
\begin{gather*}
{ I + B   ( \cos r_{0} -1 ) \over  \sin ^{2} r_{0} }  = -   { B
\over \cos  r_{0}}  \qquad \Longrightarrow \qquad I = B   {\cos
r_{0}  -1  \over  \cos  r_{0}  }   ,
\end{gather*}
substituting it into  (\ref{10.13}) we arrive at
\begin{gather*}
A =     { [  I + B   ( \cos  r_{0} -1)  ]^{2} \over
  \sin^{2} r_{0} }
  = { B^{2} \over  \sin^{2} r_{0} }
  \left[  {(\cos  r_{0} -1 ) \over  \cos r_{0} } + (\cos  r_{0} -1 )  \right]^{2} =
   \big( \tan^{2} r_{0}    B^{2}  \big)   ,
\end{gather*}
which coincides with  (\ref{10.14}).

\section[All  trajectories  and $SO(4)$ symmetry of the space $S_{3}$]{All  trajectories  and $\boldsymbol{SO(4)}$ symmetry of the space $\boldsymbol{S_{3}}$}\label{section13}

Now,  using  general relations ({\ref{10.13a})--(\ref{10.13e}),
let us examine the general case of possible solutions. First, let
$A  \neq \epsilon $. Relation (\ref{10.13c}) is integrated
straightforwardly
\begin{gather*}
 \sin   z = \pm  {  \sqrt{\epsilon - A } \over \sqrt{\epsilon}}
 \sin  \sqrt{\epsilon} t  . %\label{11.2}
\end{gather*}
Equation  (\ref{10.13b}) reads
\begin{gather*}
\int   {  \sin  r   d r \over  \sqrt   { A   \sin^{2} r  -
  (I + B   \cos  r - B)^{2}  } } = \pm
  \int {d t \over \cos ^{2} z }     .
 %\label{11.3}
 \end{gather*}
The integral in the right-hand side is known, see
(\ref{8.8a}),
\begin{gather*}
 R =  \pm \int {d t \over  \cos ^{2} z }
 = \pm   {  1  \over \sqrt{ A} } \arctan \left( \sqrt{{A \over \epsilon}}
\tan  \sqrt{\epsilon}  t\right)   . %\label{11.4}
\end{gather*}
The integral in the left-hand side is\footnote{In should be stressed that in the
spherical model $S_{3}$ always $a<0$ which means that here only
f\/inite  motions are possible.}
\begin{gather}
L = \int   {  \sin  r   d r \over  \sqrt   { A   ( 1 - \cos^{2}
r ) -
  (I + B   \cos  r - B)^{2}  } }
 =  - \int {  dx \over
\sqrt{ a   \big(x +{b\over 2a}\big)^{2} + {b^{2}-4ac \over -4a}}}  ,
\nonumber
\\
x = \cos r  , \qquad a = -A -  B^{2} < 0  , \qquad
b =  - 2B (I - B)   , \qquad c = A - (I-B)^{2}     . \label{11.5b}
\end{gather}

\centerline{\includegraphics{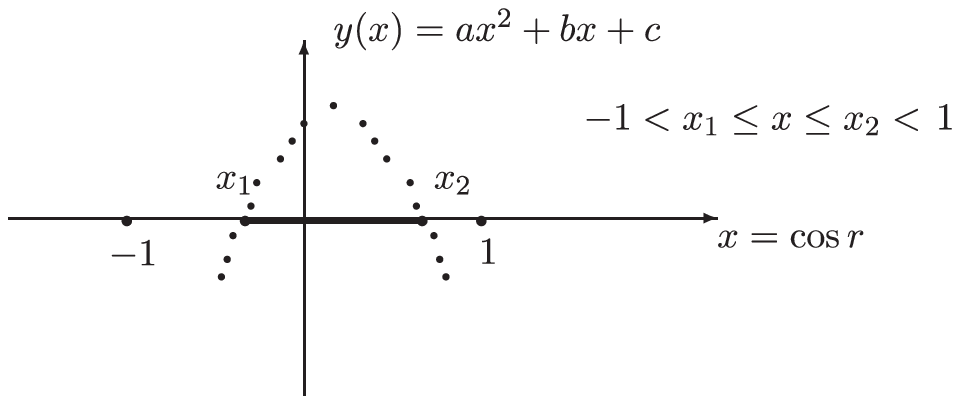}}

\centerline{{\bf Fig.  3.} Finite  motion.}

\medskip

%\vspace{-5mm}
%\unitlength=0.6 mm
%\begin{picture}(160,100)(-40,-60)
%\special{em:linewidth 0.4pt} \linethickness{0.4pt}

%\put(-20,0){\vector(+1,0){120}}     \put(+100,-5){$x= \cos  r $}
%\put(+30,-30,0){\vector(0,+1){60}}   \put(+35,+30){$y(x) = ax^{2}
%+ bx + c   $}

%\put(35,+19){\circle*{1}} \put(30,+16){\circle*{1}}
%\put(40,+16){\circle*{1}} \put(28,+13){\circle*{1}}
%\put(42,+13){\circle*{1}} \put(26,+10){\circle*{1}}
%\put(44,+10){\circle*{1}} \put(22,+6){\circle*{1}}
%\put(48,+6){\circle*{1}} \put(21,+2){\circle*{1}}
%\put(49,+2){\circle*{1}} \put(18,-3){\circle*{1}}
%\put(52,-3){\circle*{1}} \put(17,-7){\circle*{1}}
%\put(53,-7){\circle*{1}} \put(16,-10){\circle*{1}}
%\put(54,-10){\circle*{1}}
%
%
%\put(0,-0.5){\circle*{2}}  \put(-3,-8){$-1$}
%\put(60,-0.5){\circle*{2}}  \put(+60,-8){$+1$}
%
%
%\put(20,-0.5){\circle*{2}}  \put(15,+5){$x_{1}$}
%\put(50,-0.5){\circle*{2}}  \put(52,+5){$x_{2}$}

%\put(20,-0.3){\line(+1,0){30} } \put(20,-0.5){\line(+1,0){30} }
%\put(20,-0.8){\line(+1,0){30} }
%
%\put(+90,+15){$ -1 < x_{1} \leq  x  \leq x_{2} < +1 \;\;\;\;\;   $
%}
%
%\end{picture}

We may expect physical solutions when
\begin{gather*}
a < 0  , \qquad  b^{2}-4ac   > 0    , \qquad
-1 \leq {b - \sqrt{b^{2}-4ac} \over -2a} = x_{1}  < x <  x_{2} =
{b + \sqrt{b^{2}-4ac} \over -2a} \leq   1  . %\label{11.6a}
\end{gather*}
First of all, we  require
$
b^{2} - 4ac = 4A   [   A+B^{2} - (I-B)^{2}   ] >  0$ ,
 that is
\begin{gather}
b^{2} - 4ac > 0 \qquad \Longrightarrow \qquad   A+B^{2} >
(I-B)^{2}   . \label{B1}
\end{gather}
Also, we  require
\begin{gather}
  -1 \leq x_{1} < x_{2} \leq +1\qquad \Longrightarrow \qquad
  b - 2a \geq   +   \sqrt{b^{2}-4ac}     , \qquad
   b + 2a \leq -   \sqrt{b^{2}-4ac}       .\!\!\!
\label{B2}
\end{gather}
From the f\/irst and second inequalities, it follows
respectively:
\begin{gather*}
b-2a \geq 0   , \qquad a + c - b \leq 0 , \qquad
b + 2a \leq  0   , \qquad d a +c + b \leq  0    ;
\end{gather*}
they are equivalent to
\begin{alignat*}{4}
& b-2a \geq  0   ,   \qquad  && -(I - 2B)^{2} \leq 0,      \qquad &&
-1 \leq x_{1}    , &
\\
& b + 2a \leq 0   , \qquad  &&  - I^{2}  \leq  0, \qquad &&
 x_{2} \leq +1    . &
\end{alignat*}
Therefore, (\ref{B2})  will be satisf\/ied if
\begin{gather}
2a \leq  b \leq  -2a \qquad \Longrightarrow \qquad -(A +B^{2})
\leq  -B(I-B)  \leq  (A +B^{2})   . \label{B2'}
\end{gather}

 Relationships (\ref{B1} ) and (\ref{B2'}) provide us  with the constrains on parameters $(A, I, B)$
ensuring the existence of solutions in spherical space $S_{3}$.
The integral (\ref{11.5b})  reduces to
\begin{gather*}
L =  \int {- dx \over \sqrt{ a   (x +{b\over 2a})^{2} +
{b^{2}-4ac \over -4a}}}=
  {1 \over \sqrt{-a}} \arcsin { 2a \cos r  + b  \over
\sqrt{b^{2} - 4ac}}   . %\label{11.10}
\end{gather*}
Therefore, equation  (\ref{10.13b})  results in
\begin{gather*}
 { 2a  \cos r  + b  \over
\sqrt{b^{2} - 4ac}} =
 \sin   \left [ {  \sqrt{-a}   \over \sqrt{ A} } \,  \mbox{arccoth}\left( \sqrt{{A \over \epsilon}}
\tanh   \sqrt{\epsilon}  t \right)   + \Lambda   \right ]
  .
%\label{11.11b}
\end{gather*}

Now  let us consider the trajectory equation  (\ref{10.13d}) in  the
form  $F(r,z) =0 $. Its right-hand side after integration gives
\begin{gather*}
 \pm   \int   {1 \over \cos  z   }
   { dz  \over \sqrt{ \epsilon  \cos ^{2} z - A}}  =
  \pm {1 \over \sqrt{A}}  \arcsin  \sqrt{{A \over \epsilon-A}}  \tan z  .
%\label{11.12}
\end{gather*}
Therefore, the trajectory equation  (\ref{10.13d})  is
\begin{gather*}
{1 \over \sqrt{-a}} \arcsin { 2a \cos  r  + b  \over
\sqrt{b^{2} - 4ac}}  - \Lambda =
  \pm {1 \over \sqrt{A}}  \arcsin \sqrt{{A \over \epsilon-A}}  \tan z  .
%\label{11.13}
\end{gather*}

Now,  let us consider equation (\ref{10.13e})  in the form  $ F(r,
\phi)= 0 $
\begin{gather}
\int { [   I + B( \cos  r -1)   ]   dr \over \sin  r  \sqrt{
A  \sin^{2}r - [   I + B( \cos r -1)  ]^{2} }} =  \phi  .
\label{11.14a}
\end{gather}

Let us introduce a new variable
\begin{gather*}
u = { (I-B) \cos r + B \over  \sin r }   ,
\end{gather*}
the integral in (\ref{11.14a}) takes the form
\begin{gather*}
L = - \int {du \over \sqrt{ ( A  + B^{2} ) - (I-B)^{2}   - u^{2}
} } = \arccos {(I-B) \cos r + B
 \over \sin r
\sqrt{( A  + B^{2} ) - (I-B)^{2} }}    . %\label{B6}
\end{gather*}

Therefore,   the general trajectory equation $F(r, \phi)=0$ in the
model $S_{3}$    looks
\begin{gather}
  (B-I) \cos r +  \sqrt{( A  + B^{2} ) - (I-B)^{2} }   \sin r   \cos \phi = B   .
\label{B7}
\end{gather}

Let us consider the behavior of this equation with respect to
(Euclidean) shifts (0--1) in spherical space. To this end, let us
introduce two coordinate systems in $S_{3}$
\begin{alignat*}{5}
& u_{1} = \cos  z   \sin  r \cos \phi   , \qquad && u_{2} = \cos  z
  \sin  r \sin \phi   , \qquad && u_{3} = \sin  z   , \qquad  &&
u_{0} = \cos   z   \cos r  ,  &
\\
& u_{1} ' = \cos  z'   \sin  r' \cos \phi'   , \qquad && u_{2}' =
\cos  z'   \sin  r'  \sin \phi '   , \qquad && u_{3}' = \sin  z '
, \qquad  && u_{0}' = \cos   z'   \cos  r'  , &
\end{alignat*}
related by the shift
\begin{gather}
\left | \begin{array}{c}
u'_{0} \\
u'_{1} \\
u'_{2} \\
u'_{3}
\end{array} \right | =
\left | \begin{array}{cccc}
\cos  \alpha  &  \sin  \alpha &  0  &  0 \\
- \sin  \alpha  &  \cos  \alpha & 0 & 0 \\
0  &  0  &  1  &  0 \\
0  &  0  &  0  &  1
\end{array} \right |
\left | \begin{array}{c}
u_{0} \\
u_{1} \\
u_{2} \\
u_{3}
\end{array} \right | .
\label{11.18a}
\end{gather}
Equation (\ref{11.18a}) gives the relation between two
cylindric coordinate systems
\begin{gather*}
z' = z   ,  \qquad  \sin  r'  \sin \phi ' =  \sin  r  \sin
\phi   , %\label{11.18b}
\\
\sin r' \cos \phi' = -  \sin \alpha   \cos r  + \cos \alpha
\sin  r  \cos \phi   , \qquad
\cos  r' = \cos \alpha  \cos  r + \sin \alpha  \sin r \cos
\phi   , \nonumber
\end{gather*}
and inverse ones
\begin{gather*}
z = z'   ,  \qquad  \sin  r  \sin \phi  =  \sin  r'  \sin
\phi'   , %\label{11.18b'}
\\
\sin r \cos \phi =   \sin \alpha   \cos r'  + \cos \alpha
\sin  r '    \cos \phi'   , \qquad
\cos  r = \cos \alpha  \cos  r'  - \sin \alpha   \sin r'
\cos \phi '  . \nonumber
\end{gather*}

Let us transform equation (\ref{B7}) to shifted coordinate $(r',
\phi')$
\begin{gather*}
  (B-I)  [   \cos \alpha  \cos  r'  - \sin \alpha    \sin r'   \cos \phi '   ]
  \\
 \qquad{} +  \sqrt{( A  + B^{2} )+ (I-B)^{2} }   [   \sin \alpha    \cos r'  + \cos \alpha   \sin  r '    \cos \phi'
    ] = B   .
\end{gather*}
After elementary regrouping   it reads
\begin{gather}
\Big[  \cos \alpha    (B-I)    +  \sin \alpha      \sqrt{( A  +
B^{2} ) - (I-B)^{2} }    \Big]  \cos r'  \nonumber
\\
\qquad{}+  \Big[     - \sin \alpha     (B-I)    +
   \cos \alpha   \sqrt{( A  + B^{2} ) - (I-B)^{2} }  \Big ]    \sin  r '    \cos \phi'
  = B   .
\label{B9}
\end{gather}

Comparing (\ref{B9}})  with (\ref{B7}),  we  see the invariance
property of the trajectory equation
 if
the parameters are transformed according to Euclidean rotation
\begin{gather*}
B '- I ' =  \cos \alpha \;   (B-I)    +  \sin \alpha
\sqrt{( A  + B^{2} ) - (I-B)^{2} }   ,
\\
\sqrt{( A'  + B^{2} ) - (I'-B)^{2} } =
 - \sin \alpha    (B-I)    +
   \cos \alpha   \sqrt{( A  + B^{2} ) - (I-B)^{2} }  ,
 %\nonumber
 %  \label{B10}
   \end{gather*}
with notation
$
B-I = J$, $C = \sqrt{( A  + B^{2} ) - (I-B)^{2} }$,
 the trajectory equation has the following invariant form
\begin{gather}
  J \cos r +  C  \sin r  \cos \phi = B \qquad \Longrightarrow
\qquad  J'  \cos r '+  C'   \sin r'  \cos \phi' = B   ,
\label{B11}
\end{gather}
with respect to Euclidean shifts (0--1) in $S_{3}$
parameters $J$, $C$ transform according to
\begin{gather*}
J ' =  J \cos \alpha        +  C \sin \alpha         , \qquad
C'  =
 - J  \sin \alpha       +
   C \cos \alpha     .
   %\label{B12}
   \end{gather*}
   This parametric shift generated  by  symmetry of the system leaves invariant the following
   (Eucli\-dean)  combination of two parameters
\begin{gather*}
\mbox{inv} = J^{2} + C^{2} =  J^{\prime 2} + C^{\prime 2}  = A + B^{2}  \qquad
\Longrightarrow \qquad A = A' =  \mbox{inv} . %\label{B13}
\end{gather*}

By a special choice of a shift one can translate the above
equation (\ref{B11})  to more simple  forms: for instance, to
\begin{gather*}
J_{0} = \sqrt{A +B^{2}}   , \qquad  C_{0} = 0  \qquad
\Longrightarrow \qquad J_{0} \cos r_{0}   = B  ; \nonumber
\\
J_{0} = 0 , \qquad C_{0} = \sqrt{A + B^{2}}
 \qquad  \Longrightarrow \qquad   C _{0}   \sin r   \cos \phi = B   .
%\label{B14}
\end{gather*}

\section[Space  shifts  in  space $S_{3}$   and gauge symmetry in  $S_{3}$]{Space  shifts  in  space $\boldsymbol{S_{3}}$   and gauge symmetry in  $\boldsymbol{S_{3}}$}

Let us introduce two cylindric coordinate systems
\begin{gather*}
u_{1} = \cos z   \sin  r \cos \phi   , \qquad u_{2} = \cos  z
\sin r \sin \phi   , \nonumber
\\
u_{3} = \sin  z   , \qquad  u_{0} = \cos  z   \cos r   ;
\nonumber
\\
u_{1} ' = \cos z'   \sin r' \cos \phi'   , \qquad u_{2}' = \cos
z'   \sin r'  \sin \phi '   , \nonumber
\\
u_{3}' = \sin z '  , \qquad  u_{0}' = \cos  z'   \cos r'  ,
%\label{13.1b}
\end{gather*}
related by the shift  (0--1)
\begin{gather*}
\left | \begin{array}{c}
u'_{0} \\
u'_{1} \\
u'_{2} \\
u'_{3}
\end{array} \right | =
\left | \begin{array}{cccc}
\cos \alpha  &  \sin  \alpha &  0  &  0 \\
- \sin \alpha &  \cos \alpha & 0 & 0 \\
0  &  0  &  1  &  0 \\
0  &  0  &  0  &  1
\end{array} \right |
\left | \begin{array}{c}
u_{0} \\
u_{1} \\
u_{2} \\
u_{3}
\end{array} \right | , \qquad
z' = z  , \qquad \sin r'  \sin \phi ' =  \sin  r  \sin \phi   ,
\nonumber
\\
\sin  r'  \cos \phi' = - \sin  \alpha  \cos r + \cos  \alpha  \sin
r  \cos \phi  , \qquad
\cos r' = \cos  \alpha  \cos r + \sin \alpha  \sin  r  \cos \phi
 . %\label{13.2b}
\end{gather*}

Under this change of variables,  $(r, \phi)   \Longrightarrow
(r', \phi')$, the  uniform magnetic f\/ield is transformed according to
\begin{gather*}
F_{\phi' r'} = {\partial x^{\alpha} \over \partial \phi '}
{\partial x^{\beta} \over \partial r'} F_{\alpha \beta}= \left(
{\partial \phi \over \partial \phi '}  {\partial r  \over \partial
r'} -{\partial r \over \partial \phi ' }  {\partial \phi \over
\partial r'}  \right) F_{\phi r}     , \qquad F_{\phi r} = B  \;
\sinh\; r   ,   %\label{13.3a}
\end{gather*}
or in terms of the Jacobian
\begin{gather*}
F_{\phi' r'} = J  F_{\phi r}   , \qquad J = \left |
\begin{array}{cc}
{\partial r \over \partial r' }   &  {\partial r \over \partial \phi ' }  \vspace{2mm}\\
{\partial \phi  \over \partial r' }  & {\partial \phi  \over
\partial \phi' }
\end{array} \right |   , \qquad   F_{\phi r}  = B    \sin r   .
%\label{13.3b}
\end{gather*}

The coordinate transformation can be presented as
\begin{gather*}
\phi =  \arctan \left (    { \sin r'   \sin \phi ' \over
 \sin  \alpha \cos  r' + \cos \alpha   \sin r' \cos \phi' } \right )
 = \arctan   A   ,
\nonumber
\\
r = \arccos \left (   \cos \alpha \cos r'  - \sin \alpha
\sin r'  \cos \phi'  \right  )= \arccos  B  .
%\label{13.4a}
\end{gather*}
The Jacobian reads
\begin{gather*}
J =  {-1 \over \sqrt{1-B^{2}}}  {1 \over 1 +A^{2}}   \left (
{\partial B \over \partial r' }   {\partial A \over \partial \phi
'} - {\partial B \over \partial \phi' }   {\partial A \over
\partial r '}   \right )   . %\label{13.4b}
\end{gather*}
With the help of an identity
\begin{gather*}
{-1 \over \sqrt{1- B^{2} }}  {1 \over 1 +A^{2}}= {-1 \over \sqrt{
1 - \cos^{2} r }} {1 \over 1 + \tan^{2} \phi}= - {\cos^{2}
\phi \over  \sin  r}  ,
\end{gather*}
 and the formulas
\begin{gather*}
{\partial B \over \partial r' } = {\partial  \over \partial r' }
 (  \cos \alpha \cos r'  - \sin \alpha \sin r'  \cos \phi' )=
- \cos \alpha  \sin  r'  - \sin \alpha \cos r'  \cos \phi'  ,
\nonumber
\\
{\partial B \over \partial \phi' } = {\partial  \over \partial
\phi ' } (  \cos \alpha \cos r'  - \sin \alpha \sin r'  \cos
\phi' )=
 \sin \alpha \sin  r' \sin \phi'  ,
\nonumber
\\
{\partial A \over \partial \phi' }= {\partial  \over \partial
\phi' }  \left (   { \sin r' \sin \phi ' \over
 \sin \alpha \cos r' + \cos \alpha  \sin r' \cos \phi' } \right )
 \nonumber
 \\
\phantom{{\partial A \over \partial \phi' }}{} =  { \sin r'   ( \sin \alpha \cos r' \cos \phi' +
 \cos  \alpha \sin  r' )  \over
( \sin \alpha \cos r' + \cos  \alpha \sin r' \cos \phi' )^{2} }= {
\sin r'   ( \sin  \alpha \cos  r' \cos \phi' +
 \cos \alpha \sin r' )  \over  \sin ^{2} r \cos^{2} \phi }  ,
\nonumber
\\
{\partial A \over \partial r' }= {\partial  \over \partial r' }
\left (   { \sin  r'  \sin \phi ' \over
 \sin  \alpha  \cos  r' + \cos  \alpha  \sin r' \cos \phi' } \right )
 \nonumber
 \\
 \phantom{{\partial A \over \partial r' }}{} =
 {  \sin  \alpha  \sin \phi '
 \over
(  \sin \alpha \cos r' + \cos \alpha  \sin r' \cos \phi' )^{2} }=
{ \sin \alpha \sin \phi '
 \over   \sin^{2} r  \cos^{2} \phi }  ,
\end{gather*}
for the Jacobian we get
\begin{gather*}
J=   { \sin r'   \over  \sin r}   { (   \cos  \alpha \sin  r'  +
\sin \alpha \cos r' \cos \phi' )
  ( \sin  \alpha \cos  r' \cos \phi' +
 \cos \alpha \sin  r' )  +
 \sin^{2} \alpha  \sin ^{2}\phi'  \over \sin ^{2} r } .
% \label{13.5c}
 \end{gather*}
From whence it follows
\begin{gather*}
J =  { \sin  r'   \over  \sin  r},
 %\label{13.6a}
\end{gather*}
therefore, a magnetic f\/ield is invariant under the shift
(0--1)  in space  $S_{3}$
\begin{gather*}
F_{\phi' r'} = J  F_{\phi r}   , \qquad
  F_{\phi r}  = B    \sin  r   , \qquad F_{\phi' r'} =  B    \sin r'  .
%\label{13.6b}
\end{gather*}

By symmetry reason, the same behavior of a magnetic f\/ield will take place
for shifts of the type  (0--2). However, for  shifts of the type
(0--3)
\begin{gather*}
\left | \begin{array}{c}
u'_{0} \\
u'_{1} \\
u'_{2} \\
u'_{3}
\end{array} \right | =
\left | \begin{array}{cccc}
\cos  \alpha  &  0 &  0  &  \sin \alpha  \\
0  &  1 & 0 & 0 \\
0  &  0  &  1  &  0 \\
-\sin \alpha   &  0  &  0  &  \cos \alpha
\end{array} \right |
\left | \begin{array}{c}
u_{0} \\
u_{1} \\
u_{2} \\
u_{3}
\end{array} \right | , \qquad
 \cos z' \cos  r' = \cos \alpha \cos  z \cos r +
 \sin  \alpha \sin z z  ,
\nonumber
\\
 \sin  z' = - \sin \alpha \cos  z \cos  r +
\cos  \alpha \sin  z   , \qquad
\cos z' \sin r' = \cos z  \sin r  , \qquad \phi ' = \phi   .
%\label{13.8a}
\end{gather*}
electromagnetic f\/ield is transformed as
\begin{gather*}
F_{\alpha ' \beta'} = \left( {\partial \phi \over \partial x^{\prime \alpha}}
{\partial r \over \partial x^{\prime \beta}} - {\partial r \over
\partial x^{\prime \alpha}}  {\partial \phi  \over \partial x^{\prime \beta}}
\right) F_{ \phi r} \qquad  \Longrightarrow \qquad
F_{\phi' r'} = {\partial r \over \partial r'}  F_{\phi r}  ,
\qquad  F_{\phi' z'} = {\partial r \over \partial z'}  F_{\phi
r} ,  %\label{13.8b}
\end{gather*}
 therefore a magnetic f\/ield is not invariant under these
shifts (0--3).

In terms of 4-potentials, the transformation rule reads
\begin{gather*}
 A_{\phi} =   B  ( \cos  r -1 ) \qquad  \Longrightarrow \qquad
 A'_{\phi'} = {\partial \phi   \over \partial \phi' }    A_{\phi}   , \qquad
A'_{r'} = {\partial \phi   \over  \partial r '}   A_{\phi}
% \label{13.9}
 \end{gather*}
and  we  may expect
\begin{gather}
 A'_{\phi'} = {\partial \phi   \over \partial \phi' }   A_{\phi}  =
  B  ( \cos  r' -1 ) +  {\partial \over \partial \phi'}   \Lambda  ,
\qquad A'_{r'} = {\partial \phi   \over  \partial r '}   A_{\phi}
=   {\partial \over \partial r '} \Lambda  .
 \label{13.11}
 \end{gather}

Accounting for two relations
\begin{gather*}
{\partial \phi \over \partial \phi '} = {1 \over 1 + A^{2} }
{\partial A \over \partial \phi '} = { \sin r'    ( \sin  \alpha
\cos  r' \cos \phi' +
 \cos  \alpha \sin  r' )  \over  \sin^{2} r  }  ,
\\
{\partial \phi \over \partial r '} = {1 \over 1 + A^{2} }
{\partial A \over \partial r '} = { \sin  \alpha   \sin \phi '
 \over   \sin^{2} r   }  ,
\end{gather*}
for a 4-potential in shifted coordinates we have
\begin{gather*}
 A'_{\phi'} = { \sin  r'   (  \cos  \alpha \sin  r' + \sin \alpha \cos  r' \cos \phi'
  )  \over  \sin ^{2} r  }    [ B ( \cos  r -1 )  ]  ,
\nonumber
\\
A'_{r'} = {\sin  \alpha  \sin \phi '
 \over   \sin^{2} r   }
   [ B ( \cos  r -1 )   ]   ,
%\label{13.12}
\end{gather*}
from whence it follows
\begin{gather*}
 A'_{\phi'} =  - B
{ \sin  r'    (  \cos  \alpha \sin  r' + \sin  \alpha \cos  r'
\cos \phi'
  )  \over
 1 + \cos \alpha  \cos  r'  - \sin \alpha \sin  r'  \cos \phi'     }   ,
\nonumber
\\
A'_{r'}  =  - B   {  \sin  \alpha  \sin \phi '
 \over    1 + \cos  \alpha \cos r'  - \sin  \alpha  \sin  r'   \cos \phi'      }   .
%\label{13.13}
\end{gather*}

The gauge function $\Lambda$  in  (\ref{13.11})  is def\/ined by its
partial derivatives
\begin{gather}
{\partial \Lambda \over \partial r'} = - B   {  \sin  \alpha \sin
\phi '
 \over    1 + \cos  \alpha \cos  r'  - \sin \alpha \sin  r'   \cos \phi'      }   ,
 \nonumber
 \\
{\partial \Lambda \over \partial \phi'} = - B (\cos  r' -1) - B\;
{ \sin  r'   (  \cos \alpha \sin  r' + \sin \alpha  \cos  r' \cos
\phi'
  )  \over
 1 + \cos  \alpha \cos  r'  - \sin \alpha  \sin  r' \; \cos \phi'     }
\nonumber
\\
\phantom{{\partial \Lambda \over \partial \phi'}}{} = -B  { ( \cos r' -1) (1 - \cos \alpha) + \sin  \alpha \sin r'
\cos \phi '\over
 1 + \cos  \alpha \cos r'  - \sin  \alpha  \sin  r'  \cos \phi'     }.
\label{13.14}
\end{gather}
Integrability condition
 in the explicit form reads
\begin{gather*}
{\partial \over \partial \phi '}  \left ( -B  {  \sin  \alpha
\sin \phi '
 \over    1 + \cos  \alpha \cos  r'  - \sin \alpha \sin  r' \cos \phi'      } \right )
\nonumber
\\
\qquad{} = {\partial \over  \partial r'}  \left ( -B
  { ( \cos  r' -1) (1 - \cos  \alpha) + \sin  \alpha \sin r' \cos \phi '\over
 1 + \cos \alpha \cos r'  - \sin  \alpha \sin  r'   \cos \phi'     } \right )   .
%\label{13.15}
\end{gather*}

Now we are to f\/ind the gauge function
 $\Lambda$. Let us integrate the f\/irst equation in
 (\ref{13.14})
\begin{gather*}
 \Lambda  = -  B   \sin \alpha  \sin \phi '  \int
   {  dr' \over    1 + \cos \alpha   \cos  r'  - \sin  \alpha
\sin r'   \cos \phi'      } + \lambda ( \phi ')   .
%\label{13.17a}
\end{gather*}
With the notation
\begin{gather*}
\cos \alpha = c  , \qquad  \sin  \alpha = s  , \qquad c^{2} +
s^{2} = 1  , \qquad
\tan  {r' \over 2} = y   , \qquad dr' = 2   \cos^{2} r'
  dy   , \\
   {  dr'
 \over    1 + \cos \alpha  \cos r'  - \sin \alpha \; \sin r' \cos  \phi'      }
\nonumber
\\
\qquad{}=  {  2 \; \cos^{2} r' \; dy
 \over    1 + c  ( \cos^{2}{ r'\over 2} - \sin^{2}{ r'\over 2} )
  - 2s     \sin { r'\over 2} \; \cos { r'\over 2}    \cos \phi'      }=
 {  2    dy
 \over    y^{2}(1-c)
  - 2 y   s   \cos \phi'    + c + 1     }  ,
%\label{13.17b}
\end{gather*}
 for  $\Lambda$  we  obtain
\begin{gather*}
\Lambda = \lambda (\phi') - 2B \arctan  { (1-c) (1 - \cos
r')  - s    \sin  r'   \cos \phi ' \over s   \sin r'     \sin
\phi '}   .
% \label{13.17c}
 \end{gather*}
Now, let us proceed further
\begin{gather*}
{\partial \Lambda \over \partial \phi '} = {d \lambda \over d \phi
'} -2B  {s^{2}  \sin^{2}r'    \sin^{2} \phi ' - [   (1-c) (1-
\cos r' )  - s    \sin r'  \cos \phi '  ]  s    \sin r' \cos
\phi'
 \over s^{2}  \sin^{2}r'     \sin^{2} \phi ' +
 [   (1-c) (1- \cos  r')  - s     \sin r'  \cos \phi '  ]^{2} }
\\
\phantom{{\partial \Lambda \over \partial \phi '}}{}
= {d \lambda \over d \phi '} - 2B  {(1+c)  (1 + \cos  r')   -
   s   \sin r' \cos \phi
 \over (1+c)   (1+ \cos r')   +
  (1-c) (1- \cos  r')  - 2 s   \sin r'   \cos \phi ' },
\end{gather*}
and f\/inally
\begin{gather*}
{\partial \Lambda \over \partial \phi '} = {d \lambda \over d \phi
'} - B  { (1+c) (1 + \cos  r')    -
   s   \sinh  r' \cos \phi
 \over
  1+ c   \cos  r'   -  s   \sin r'   \cos \phi ' }   .
 %\label{13.18a}
 \end{gather*}
Substituting it into the second equation in
 (\ref{13.14}) we get
 \begin{gather*}
 {d \lambda \over d \phi '} =
-B  { ( \cos r' -1) (1 - c) + s \sin r' \cos \phi '\over
 1 + c  \cos r'  - s  \sin  r'  \cos \phi'     }
 +
B  { (1+c) (1 + \cos  r')    -
   s   \sinh  r' \cos \phi
 \over
  1+ c  \cos  r'   -  s    \sin r'  \cos \phi ' }
  = 2B   .
  \end{gather*}
Therefore,
\begin{gather*}
\lambda (\phi)  =   2B \phi ' + \lambda _{0}  .
 %\label{12.19}
 \end{gather*}
Thus, the gauge function  $\Lambda (r', \phi')$ is
found (where  $ c = \cos \alpha$, $s = \sin \alpha $)
\begin{gather*}
\Lambda (r', \phi')= - 2B \arctan { (1-c) (1 - \cos  r')
- s   \sin  r'  \cos \phi ' \over s  \sin r'    \sin \phi
'}
 + 2B \phi' + \lambda _{0}  .
%\label{13.20}
\end{gather*}

\section{Extension to relativistic case}\label{section15}

Now let us brief\/ly consider an extension to the relativistic problem~--
particle in a magnetic f\/ield in spaces~$H_{3}$ and~$S_{3}$. It
suf\/f\/ices   to discuss  the hyperbolic case, then the relativistic
equations have the form
\begin{gather*}
 {d \over dt}
 ( { 1   \over \sqrt{1 - V^{2} / c^{2}} } ) = 0  \qquad \Longrightarrow
\qquad
 { 1   \over \sqrt{1 - V^{2} / c^{2}} } =  {1 \over  \lambda } =  \mbox{const},
    \qquad \lambda = { mc^{2} \over E } < 1
%\label{18.1a}
\end{gather*}
and
\begin{gather*}
{d \over dt} {V^{1} \over \sqrt{1 - V^{2} /c^{2}}} + {1 \over
\sqrt{1 - V^{2} /c^{2}}}   \Gamma^{1}_{\;\;jk }   {dx^{j} \over
d t } {d x^{k} \over d t} =  -  \big( V_{2} B^{3} - V_{3} B^{2}\big)
, \nonumber
\\
 {d \over dt} {V^{2} \over \sqrt{1 - V^{2} /c^{2}}} +
{1 \over \sqrt{1 - V^{2} /c^{2}}}  \Gamma^{2}_{\;\;jk }
{dx^{j} \over d t } {d x^{k} \over d t} = -  \big( V_{3} B^{1} - V_{1}
B^{3}\big)   , \nonumber
\\
 {d \over dt} {V^{3} \over \sqrt{1 - V^{2} /c^{2}}} +
{1 \over \sqrt{1 - V^{2} /c^{2}}} \; \Gamma^{3}_{\;\;jk }
{dx^{j} \over d t } {d x^{k} \over d t} = - q\big( V_{1} B^{2} - V_{2}
B^{1}\big)   % \label{18.1b}
\end{gather*}
or in the explicit form
\begin{gather*}
{d V^{r} \over d t } + 2   \tanh  z   V^{r} V^{z} -
\sinh  r   \cosh  r   V^{\phi } V^{\phi}   = \lambda
  B   { \sinh  r \over \cosh^{2} z }  V^{\phi}   ,
\nonumber
\\
{d V^{\phi} \over d t } + 2   \coth  r   V^{\phi } V^{r} +
2   \tanh  z   V^{\phi } V^{z}  =
 - \lambda   B   {1 \over \cosh^{2} z   \sinh  r}   V^{r}   ,
\nonumber
\\
 {d V^{z} \over d t } -\sinh  z  \cosh  z   V^{r} V^{r}
-\sinh  z   \cosh  z   \sinh^{2} r   V^{\phi}
V^{\phi}  = 0   . %\label{18.2}
\end{gather*}

Therefore, all calculations performed for non-relativistic case
are  valid for relativistic case as well with the only change $B
\Longrightarrow \lambda  B$ and additional restriction
$\epsilon <1$.

\section{Discussion}\label{section16}

Let us summarize the main results of the paper.

Motion of a classical  particle in 3-dimensional Lobachevsky  and
Riemann spaces  is studied in the presence of an external magnetic
f\/ield which is analogous to a constant uniform magnetic
 f\/ield in Euclidean space.
 In both cases the   equations of motion
 are solved exactly in special cylindrical coordinates.
 In Lobachevsky space there exist trajectories of two types: f\/inite and inf\/inite in radial variable,
in Riemann space all  motions are f\/inite and periodical.
 The invariance   of the uniform magnetic f\/ield in tensor description and gauge invariance of corresponding
 4-potential description is demonstrated explicitly.
The role  of the symmetry is  clarif\/ied in classif\/ication of all
possible solutions given, based on the geometric  symmetry
group, $SO(3,1)$ and $SO(4)$ respectively.

{\sloppy Several additional points  should be  mentioned.
Magnetic f\/ields introduced in the  mo\-dels~$H_{3}$,~$S_{3}$ are not
invariant under geometrical shift of the type (0--3), instead
these geometrical  transformations  generate some additional
electric f\/ields. So, instead of four  symmetry generators in Euclidean  space,
$
(P_{1}, P_{2},P_{3},J_{z})$,  in curved space
models  we have only 3 generators for symmetries, $(J_{01}, J_{02},
J_{z})$.  This means that  the magnetic f\/ields under consideration
 in curved spaces    are ``less
uniform''  than in Euclidean space.

}

The choice of special coordinate systems is  a  matter of principal importance when exploring  any problem,
 and our special choice of cylindrical coordinates is not accidental but  turns  to be decisive  one.
 For instance, in Lobachevsky and Riemann  models there exist other cylindrical coordinates in which Maxwell equations
 can be solved as well.  In this case,  electromagnetic potential is (in $H_{3}$ model)
\begin{gather*}
dS^{2} = dt^{2} - dr^{2} - dr^{2} - \sinh^{2} d \phi^{2} - \cosh^{2} d z^{2}  ,\qquad
A_{\phi} = \mbox{const}\,  [  \ln   ( \cosh   r )  ]
\nonumber
\\
u_{0} = \cosh r   \cosh  z  , \qquad
u_{3} = \cosh  r   \sinh  z   ,\qquad
u_{1} = \sinh  r   \cos \phi   , \qquad
u_{0} = \sinh  r   \sin \phi   ,
\end{gather*}
 such a potential has a good limiting behavior at vanishing curvature limit  and   admits separating of variables,
 but    it  hardly could   be brought to explicit  analytical solutions
 when dealing with a~particle behavior   in this f\/ield.

Up to now,  the most attention in the literature was given  to a Kepler problem
(quantum mechanical and classical) in Lobachevsky
and Riemann models and to the
magnetic monopole problem. Corresponding electromagnetic potentials  (in spherical coordinates) are
\begin{gather*}
H_{3}   , \qquad A_{0} = {e \over \tanh  r} ,\qquad  A_{\phi} = g \cos \theta  ; \qquad
S_{3}   , \qquad A_{0} = {e \over \tan  r}  , \qquad  A_{\phi} = g \cos \theta  .
\end{gather*}
Magnetic f\/ield potentials  used by us are
\begin{gather*}
H_{3}   , \qquad A_{\phi} = -B (\cosh\; r - 1)  ; \qquad
S_{3}   , \qquad A_{\phi} = B (\cos r - 1).
\end{gather*}
They all  provide us with solutions of the Maxwell equations.
So all three potentials are equally correct. They  are equally interesting as problems simple enough
for their  analytical treatment and as   extensions of classical physical problems
in f\/lat space.
At present time, we think,  the third one is most interesting because  one may
 expect new results on this f\/ield.
  For instance, as shown in \cite{BRK-3} there exists special
additional electric
 f\/ield that allows for solutions of corresponding Schr\"{o}dinger equations in terms of hypergeometric  functions,
 so one may  expect the respective solutions of the classical equations in the presence of these additional electric f\/ields
 as well. In this case one might expect to extend symmetry operations governing the structure
 of all solutions in presence of  both   uniform magnetic and   electric f\/ields.

\subsection*{Acknowledgements}

Authors are  grateful  to  participants of seminar of Laboratory
of Theoretical Physics,
 National Academy of Sciences of Belarus for  moral support and  advice,
also we are grateful to  anonymous   reviewers  for  stimulating discussion and criticism.
This  work was  also supported  by the Fund for Basic Researches of Belarus
 F09K-123.
 We wish to thank the Organizers of the VIII-th International Conference ``Symmetry
in Nonlinear Mathematical Physics'' (June 21--27, 2009, Kyiv)   for
having given us the opportunity to talk on this subject as well as
for local support.

\pdfbookmark[1]{References}{ref}
\LastPageEnding

\end{document}